\documentclass[notitlepage,superscriptaddress,nofootinbib,tightenlines,preprintnumbers,twocolumn]{revtex4-2}
\usepackage{amsmath,verbatim,latexsym,amssymb,indentfirst,mathrsfs,mathtools,amsthm,bbm,bm,hyperref,url,cancel,subcaption,enumitem}
\usepackage[font=small,labelfont=bf,format=plain,justification=raggedright,singlelinecheck=false]{caption}
\usepackage{verbatim,indentfirst}
\usepackage[title,titletoc]{appendix}

\usepackage{graphicx}
\usepackage{epstopdf}
\usepackage{adjustbox}
\renewcommand{\thesection}{\Roman{section}}
\renewcommand{\thesubsection}{\Roman{section} \Alph{subsection}}
\renewcommand{\thesubsubsection}{\Roman{section} \Alph{subsection} \arabic{subsubsection}}
\makeatletter
\def\p@subsection{}
\makeatother
\makeatletter
\def\p@subsubsection{}
\makeatother

\makeatletter
\newcommand\footnoteref[1]{\protected@xdef\@thefnmark{\ref{#1}}\@footnotemark}
\makeatother

\usepackage{soul}

\newcommand{\QFI}{\mathcal{I}}
\newcommand{\ps}{ {\rm ps} }
\newcommand{\prior}{{\rm p}}

\newcommand{\KD}{ \tilde{p} }

\newcommand{\Min}{ {\rm min} }   
\newcommand{\Max}{ {\rm max} }   
\newcommand{\Tr}{{\rm Tr}}   % Trace
\def\id{\mathbbm{1}}   % Identity
\newcommand{\1}{ {(1)} }
\newcommand{\2}{ {(2)} }

\newcommand{\LParen}{ \bm{(} }
\newcommand{\RParen}{ \bm{)} }

\newcommand*{\Set}[1]{\left\{  #1  \right\}}

\renewcommand\th{ {\rm th} }

\newcommand*{\ket}[1]{\lvert #1 \rangle}
\newcommand*{\braket}[2]{\langle #1 \lvert #2 \rangle}
\newcommand*{\ketbra}[2]{\lvert #1 \rangle\!\langle #2 \rvert}

\usepackage{color}
\usepackage[dvipsnames]{xcolor}
\usepackage[normalem]{ulem}

\begin{document}

% \preprint{APS/123-QED}

\title{Negative quasiprobabilities enhance phase estimation in quantum-optics experiment}

\author{Noah Lupu-Gladstein}
\email{nlupugla@physics.utoronto.ca. The first two coauthors contributed equally.}
\affiliation{CQIQC and Department of Physics, University of Toronto, 60 Saint George St., Toronto, ON M5S 1A7, Canada}
\author{Y. Batuhan Yilmaz}
\email{ybylmaz@physics.utoronto.ca.}
\affiliation{CQIQC and Department of Physics, University of Toronto, 60 Saint George St., Toronto, ON M5S 1A7, Canada}
\author{David R. M. Arvidsson-Shukur}
\email{drma2@cam.ac.uk}
\affiliation{Hitachi Cambridge Laboratory, J. J. Thomson Ave., Cambridge CB3 0HE, United Kingdom}
\affiliation{Cavendish Laboratory, Department of Physics, University of Cambridge, Cambridge CB3 0HE, United Kingdom}
\author{Aharon Brodutch}
\email{brodutch@physics.utoronto.ca}
\affiliation{CQIQC and Department of Physics, University of Toronto, 60 Saint George St., Toronto, ON M5S 1A7, Canada}
\author{Arthur O. T. Pang}
\email{arthur.pang@mail.utoronto.ca}
\affiliation{CQIQC and Department of Physics, University of Toronto, 60 Saint George St., Toronto, ON M5S 1A7, Canada}
\author{Aephraim M. Steinberg}
\email{steinberg@physics.utoronto.ca}
\affiliation{CQIQC and Department of Physics, University of Toronto, 60 Saint George St., Toronto, ON M5S 1A7, Canada}
\author{Nicole Yunger Halpern}
% \email{nicoleyh@umd.edu}
\affiliation{Joint Center for Quantum Information and Computer Science, NIST and University of Maryland, College Park, MD 20742, USA}
\affiliation{Institute for Physical Science and Technology, University of Maryland, College Park, MD 20742, USA}
\affiliation{ITAMP, Harvard-Smithsonian Center for Astrophysics, Cambridge, MA 02138, USA}
\affiliation{Department of Physics, Harvard University, Cambridge, MA 02138, USA}
% \affiliation{Research Laboratory of Electronics and Center for Theoretical Physics, Massachusetts Institute of Technology, Cambridge, Massachusetts 02139, USA}

\date{\today}

\begin{abstract}

% < 200 words
Operator noncommutation, a hallmark of quantum theory, limits measurement precision, according to uncertainty principles. Wielded correctly, though, noncommutation can boost precision. A recent foundational result relates a metrological advantage with negative quasiprobabilities---quantum extensions of probabilities---engendered by noncommuting operators. We crystallize the relationship in an equation that we prove theoretically and observe experimentally. Our proof-of-principle optical experiment features a filtering technique that we term \emph{partially postselected amplification} (PPA). Using PPA, we measure a waveplate's birefringent phase. PPA amplifies, by over two orders of magnitude, the information obtained about the phase per detected photon. In principle, PPA can boost the information obtained from the average filtered photon by an arbitrarily large factor. The filter's amplification of systematic errors, we find, bounds the theoretically unlimited advantage in practice. PPA can facilitate any phase measurement and mitigates challenges that scale with trial number, such as proportional noise and detector saturation. By quantifying PPA's metrological advantage with quasiprobabilities, we reveal deep connections between quantum foundations and precision measurement.
\end{abstract}
\maketitle

\emph{Introduction.---}Advances in quantum metrology have engendered new measurement techniques \cite{Rozema12-2, giovannetti2004quantum, Giovanetti06, Giovanetti11, Polino20}. The paradigmatic quantum measurement is phase estimation, whose applications span polarimetry, magnetic sensing, gravitational-wave astronomy, and quantum-computer calibration~\cite{yoon2020experimental, Abadie2011, Abbott16, ghosh2011, budker2007optical, knill2008randomized, dobvsivcek2007arbitrary}.
A fundamental limit bounds how precisely one can estimate a phase from a given number of trials~\cite{Cramer16, Rao92}. 
If some trials are filtered out, the average information per retained, or \emph{postselected}, trial can exceed this limit~\cite{DRMAS_20_Quantum}. Filtering can never increase the information per \emph{input} trial, so successful postselections' rarity counterbalances the extra information~\cite{Combes14, Ferrie14-2}.
Nevertheless, distilling information from many input trials into fewer postselected trials can alleviate challenges that scale with trial number, including detector saturation, proportional noise, low-frequency noise, limited memory, and limited computational power~\cite{Harris17, Sinclair17, Hallaji_17_Weak, Viza15, Qiu17, Harris17}.

We elucidate this distillation's physical and mathematical roots
using a filtering technique that we call \emph{partially postselected amplification} (PPA). Theoretically, the information obtained per PPA trial can diverge as the fraction of postselected trials vanishes~\cite{DRMAS_20_Quantum}.
A related technique, weak-value amplification, offers a similarly diverging advantage~\cite{Vaidman88, Duck89, Hosten08, Dixon09, Pang14, Pang15, Jordan14, Starling09, Starling10, Magana14, Lyons14, Martinez17, Egan12, Harris17, Hallaji_17_Weak, Hofmann12, Qiu17, Viza15}.
Both techniques are examples of \emph{noncommutative filtering}.
We define \emph{noncommutative filtering} as any filtering whose effect does not depend on when the filter acts.
% We define \emph{(non)commutative filtering} as any filtering whose effect is not (in)dependent of when the filter acts.
% NYH: We say "independent of" but "dependent on."
During the alternative, \emph{commutative filtering}, the per--postselected-trial precision cannot exceed the per--input-trial limit~\cite{DRMAS_20_Quantum}. 
Examples include the neutral-density filter that reduces a camera's overexposure.
% and (ii) the chopper and lock-in amplifier that block signals outside a desired frequency band. 
PPA's postselected trials break the per--input-trial limit
by endowing a certain quasiprobability distribution with negative elements~\cite{DRMAS_20_Quantum}.

\emph{Quasiprobabilities} represent quantum states as probability densities represent states in classical statistical mechanics.
Like probabilities, the quasiprobabilities in a distribution sum to one.
Yet quasiprobabilities can assume negative and nonreal values,
called \emph{nonclassical values}. 
They can arise when the quasiprobability describes quantum-incompatible operations or observables. Well-known quasiprobability distributions include the Wigner function. 
A rising star is the \emph{Kirkwood-Dirac distribution}~\cite{Kirkwood33,Dirac45}, which has recently found applications in quantum state tomography~\cite{Johansen07, Lundeen11, Lundeen12, Bamber14, Thekkadath16}, chaos~\cite{NYH_17_Jaryznski, NYH_18_Quasiprobability, JRGA_19_Out, NYH_19_Entropic, Razieh19}, 
postselected metrology~\cite{Steinberg95,  Starling09, Dressel14, Pusey14, Pang14, Pang15,ArvShukur17-2, ArvShukur19, Kunjwal19, DRMAS_20_Quantum, Jenne21},
measurement disturbance~\cite{Jozsa07, Hofmann11, Dressel12, Monroe_21_Weak}, 
quantum thermodynamics~\cite{Allahverdyan_14_Nonequilibrium,NYH_17_Jaryznski, Miller_17_Time,Levy19, Lostaglio20}, 
and quantum foundations~\cite{Griffiths84, Goldstein95, Hartle04, Hofmann11, Hofmann12, Hofmann12-2, Hofmann14, Hofmann15, Hofmann16, Halliwell16, Stacey19,DRMAS_21_Conditions,DeBievre_21_Kirkwood,Oszmaniec_21_Measuring}. Negative Kirkwood-Dirac quasiprobabilities have been demonstrated, under certain conditions, to underlie operational advantages in quantum computation, work extraction, and parameter estimation~\cite{DRMAS_20_Quantum,JRGA_19_Out,Lostaglio20,Jenne21}.

We demonstrate PPA's parameter-estimation enhancement 
in a proof-of-principle polarimetry experiment. We estimate the birefringent phase imparted to photons by a near-half--waveplate. A tunable polarization filter implements the PPA. The filter boosts the per--detected-photon precision by over two orders of magnitude.
Furthermore, we measure a Kirkwood-Dirac distribution that describes the experiment. Our experiment operationally motivates a measure of the distribution's negativity.
We prove theoretically and confirm experimentally that the negativity is proportional to the precision enhancement when the phase is probed optimally.
We also pinpoint which systematic errors limit
PPA's theoretically unbounded precision enhancement (App. ~\ref{app_systematic}). 
Our experiment unifies theoretical quantum foundations with practical precision measurement.

\begin{figure*}[ht]
    \centering
    \includegraphics[width=0.9\textwidth]{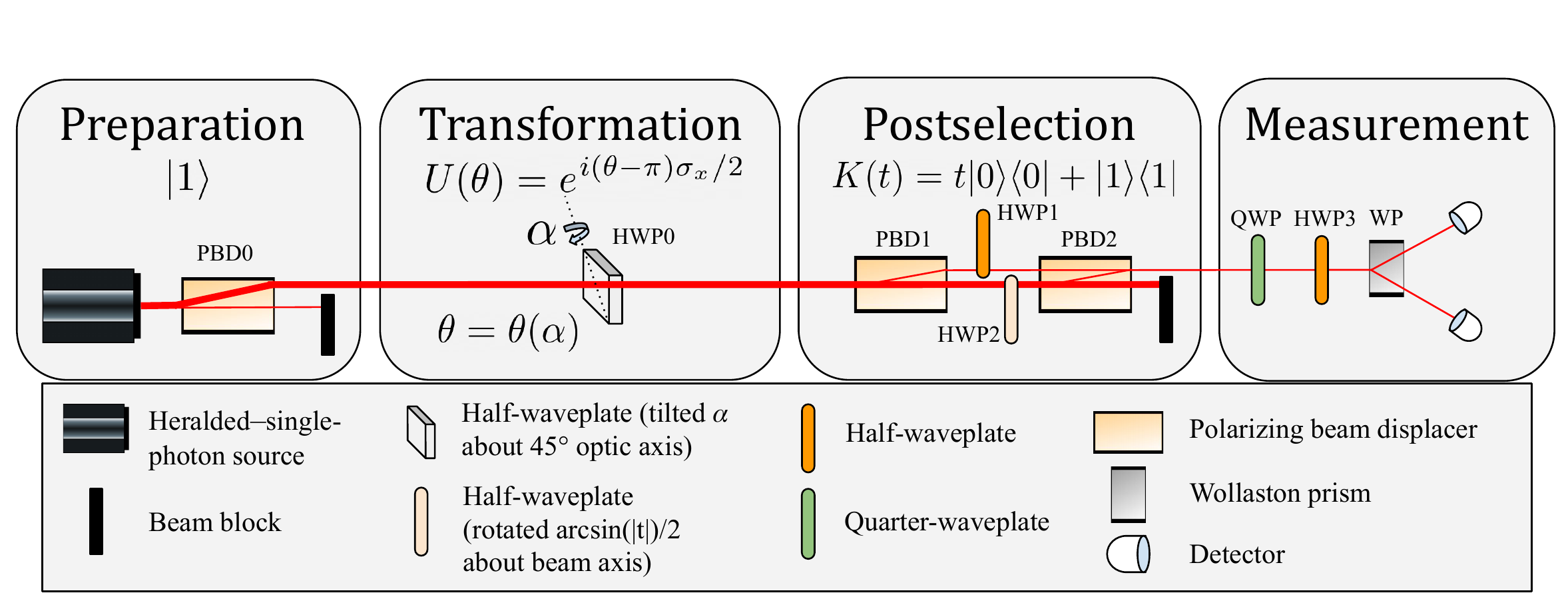}
    \caption{\textbf{Photonic parameter-estimation experiment:}
    Preparation: A heralded--single-photon source emits vertically polarized ($\ket{1}$) light. 
    Transformation: The half-waveplate ($\textrm{HWP}0$) has an optic axis angled $45^\circ$ above the horizontal. $\textrm{HWP}0$ is tilted away from normal incidence through an angle $\alpha$ about its optic axis. The waveplate rotates a photon's polarization through an angle $\theta(\alpha) - \pi$. A calibration curve of $\theta(\alpha) \equiv \theta$ provides a prior estimate of $\theta$. We use this estimate to calculate the polarization projection optimal for inferring $\theta$ (App.~\ref{app_optimal}).
    Postselection: A polarizing--beam-displacer interferometer, followed by a beam block in the undisplaced port, realizes a partial polarizer. The horizontal-polarization transmission amplitude, $t$ with $|t| \in [0, 1]$, is controlled by a half-waveplate ($\textrm{HWP}2$) inside the interferometer. The filter discards all horizontally polarized photons when $|t| = 0$ and none when $|t| = 1$.
    Measurement: Motorized waveplates, followed by a Wollaston prism ($\textrm{WP}$) and single-photon counters, project onto any desired polarization.
    }
    \label{fig_Exp_Setup}
\end{figure*}

\emph{Theoretical background and equality.---}Consider estimating a parameter $\theta$ by measuring a quantum state $\rho(\theta)$. The \emph{quantum Fisher information} (QFI) $\QFI(\theta)$ quantifies the information provided by $\rho(\theta)$ about $\theta$, via the state's sensitivity to changes in $\theta$~\cite{Braunstein94} (App.~\ref{app_optimal}).
The QFI's reciprocal lower-bounds the variance of every unbiased estimator $\theta_{\mathrm{e}}$ of $\theta$,
in the Cram\'er-Rao bound,
$\mathrm{var}(\theta_{\mathrm{e}}) \geq 1/\QFI(\theta)$.~\cite{Cramer16, Rao92}

Let $A$ denote an observable with greatest and least eigenvalues $a_+$ and $a_- = a_+ - \Delta$. The eigenstates $\ket{a_\pm}$ satisfy $A \ket{a_\pm} = a_\pm \ket{a_\pm}$.
Let a unitary $U(\theta) = \exp(i \theta A)$ imprint $\theta$ on an input state. 
The optimal inputs are even-weight superpositions of extremal $A$ eigenstates, e.g., 
$ \ket{0} = ( \ket{a_+} + \ket{a_-} ) / \sqrt{2}$ and 
$\ket{1} = (\ket{a_+} - \ket{a_-}) / \sqrt{2}$. 
The imprinted state
$U(\theta) \ket{0} = \ket{\Psi(\theta)}$
carries the most QFI possible without postselection, 
$\QFI(\theta) = \Delta^2$.

A postselected state can provide more QFI.
If the angle is small ($\theta \Delta \ll 1$), then
$\ket{\Psi(\theta)} 
\approx \ket{0} + i \frac{\theta \Delta }{2} \ket{1}$. 
The $\ket{0}$ coefficient is less sensitive to $\theta$ than the $\ket{1}$ coefficient, yet $\ket{0}$ has a greater population. PPA partially postselects on $\ket{1}$ via
a filter whose $\ket{1}$ transmission amplitude is unity
and whose $\ket{0}$ transmission amplitude is parametrically smaller.

More precisely, let $t$ denote the amplitude for $\ket{0}$'s survival of the filter. The filter acts as the Kraus operator~\cite{Nielsen11}
$K(t) = t \ketbra{0}{0} + \ketbra{1}{1}$, wherein $|t| \in [0, 1]$.
The filter lets $\ket{\Psi(\theta)}$ pass with a probability
\begin{align}
  p^\ps(\theta, t)
  % % %
  & =  \Tr \LParen K(t) \ketbra{ \Psi(\theta) }{ \Psi(\theta) } 
  K(t)^\dag  \RParen \\
  & = |t|^2  \cos^2 (\Delta \theta / 2)
  + \sin^2 (\Delta \theta / 2).
\end{align}
The state becomes
\begin{align}
  \ket{ \Psi^\ps(\theta, t) }
  % % %
  & = K(t) \ket{ \Psi(\theta) } / \sqrt{p^\ps(\theta, t)} \\
  % % %
  \label{eq_Psi_ps}
  & =  \cos (\Delta \Theta / 2) \ket{0}
  + i \,  \sin (\Delta \Theta / 2)  \ket{1}  
    \, .
\end{align}
The filter effectively amplifies $\theta$ to 
a $\Theta$ defined through 
$\tan(\Delta \Theta / 2) = \tan(\Delta \theta / 2) / |t|$.
The postselected state carries the QFI
\begin{equation}
\label{eq_Fisher_info_theory}
    \mathcal{I}(\theta) 
    = \left[ \Delta \  |t| / p^\ps(\theta, t) \right]^2.
\end{equation}

A large angle is typically easier to observe than a smaller one. If the angle is small, $\Delta \theta \ll 1$, then $\Theta$ exceeds $\theta$ by % the amplification  <-- NYH: "a" is shorter, and we've already said that the factor makes Theta bigger than theta.
a factor $1/|t|$. 
This amplification boosts the information obtained per detected state:
$\mathcal{I}(\theta) 
\approx (\Delta / |t|)^2$.  
The amplification is arbitrarily large 
if $\Delta \theta$ is arbitrarily small. 
Such extreme filtering does not significantly reduce the information obtainable per input state:
$p^\ps (\theta, t) \mathcal{I}(\theta) 
\approx \Delta^2$, if 
$\tan(\Delta \theta/2) \ll |t|$.

PPA can be beneficial even if $\Delta \theta$ is large. 
Suppose prior knowledge indicates that 
$\theta \approx \theta_\prior \in [0, 2\pi)$. 
Performing $U(-\theta_\prior)$ after $U(\theta)$ shrinks 
the probed angle to $\Delta ( \theta - \theta_\prior)$.

\begin{table}
\centering
    \begin{tabular}{ c||c|c } 
        $\KD_{\rho(\theta), t}(a, a' | +)$ & $a' = a_+$ & $a' = a_-$ \\
        \hline
        \hline
        $a = a_+$ & $\frac{1 + |t|^2}{4 p^{\ps}(\theta, t)}$ & $e^{i \Delta \theta} \frac{-1 + |t|^2}{4 p^{\ps}(\theta, t)}$ \\
        \hline
        $a = a_-$ & $e^{-i \Delta \theta} \frac{-1 + |t|^2}{4 p^{\ps}(\theta, t)}$ & $\frac{1 + |t|^2}{4 p^{\ps}(\theta, t)}$
    \end{tabular}
\caption{Conditional Kirkwood-Dirac distribution~\eqref{eq_conditional_quasi} for our PPA experiment
and $\rho(\theta) = \ketbra{\Psi(\theta)}{\Psi(\theta)}$.}
\label{tbl_kd_dist}
\end{table}

Why can a successful PPA trial offer more information than $\Delta^2$, the most information offered by any input trial?
Reference~\cite{DRMAS_20_Quantum} identified a necessary condition.
% A necessary condition identified in~\cite{DRMAS_20_Quantum} provides a clue. 
A projectively postselected trial can carry information 
$> \Delta^2$ only if a Kirkwood-Dirac distribution contains a negative quasiprobability. We generalize that result beyond projective postselection.

Let $\{\ket{a}\}_a$ and $\{\ket{a'}\}_{a'}$ denote copies of an $A$ eigenbasis.
Kraus operators $\{K_f\}_f$ with $\sum_f K_f^\dagger K_f = \id$ model the partial postselection.
The information-bearing state $\rho(\theta)$ is represented by the Kirkwood-Dirac quasiprobabilities (App.~\ref{app_KD})
\begin{align}
  \label{eq_Gen_Quasi}
  \KD_{\rho(\theta)} (a, f, a')
  % % %
  & := \Tr \LParen | a' \rangle \langle a' | 
  K_f^\dag K_f  | a \rangle \langle a |  
%   \nonumber \\ & \qquad \quad \times
  \rho(\theta) \RParen.
\end{align}
Conditioning on a postselection outcome $f$ induces the \emph{conditional Kirkwood-Dirac distribution}
\begin{equation}
    \label{eq_conditional_quasi}
    \KD_{\rho(\theta)} (a, a' | f) 
    := \KD_{\rho(\theta)} (a, f, a') / \sum_{a, a'} \KD_{\rho(\theta)} (a, f, a').
\end{equation}
These quasiprobabilities are positive if $A$ and $K_f^\dagger K_f$ commute on the support of $\rho(\theta)$~\cite{DRMAS_21_Conditions}.

\begin{figure*}[ht]
    \centering
    \begin{subfigure}[b]{0.33\textwidth}
        \centering
        \includegraphics[width=\textwidth]{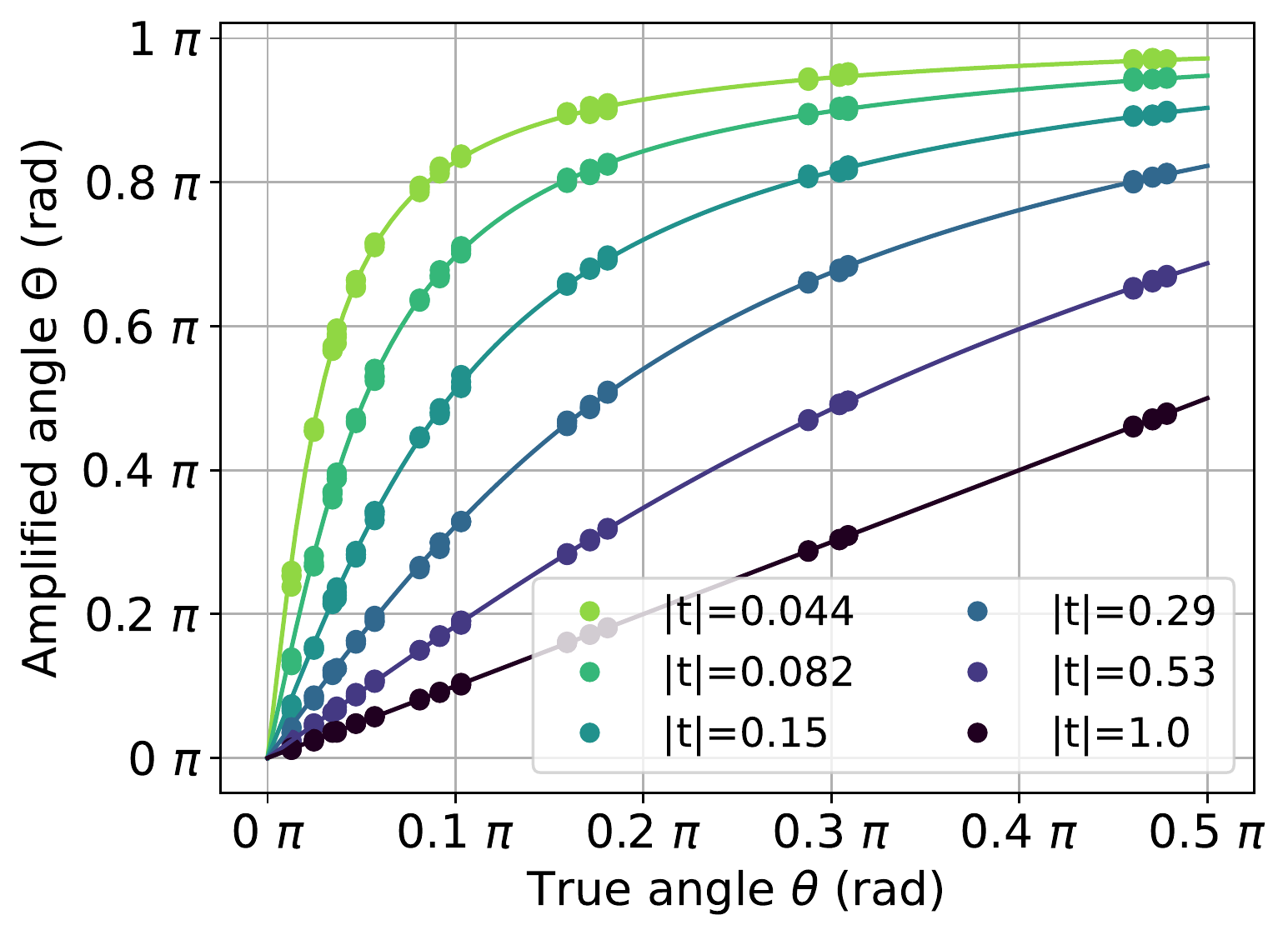}
        \caption{ }
        \label{fig_amplified_vs_original}
    \end{subfigure}
    \begin{subfigure}[b]{0.32\textwidth}
        \centering
        \includegraphics[width=\textwidth]{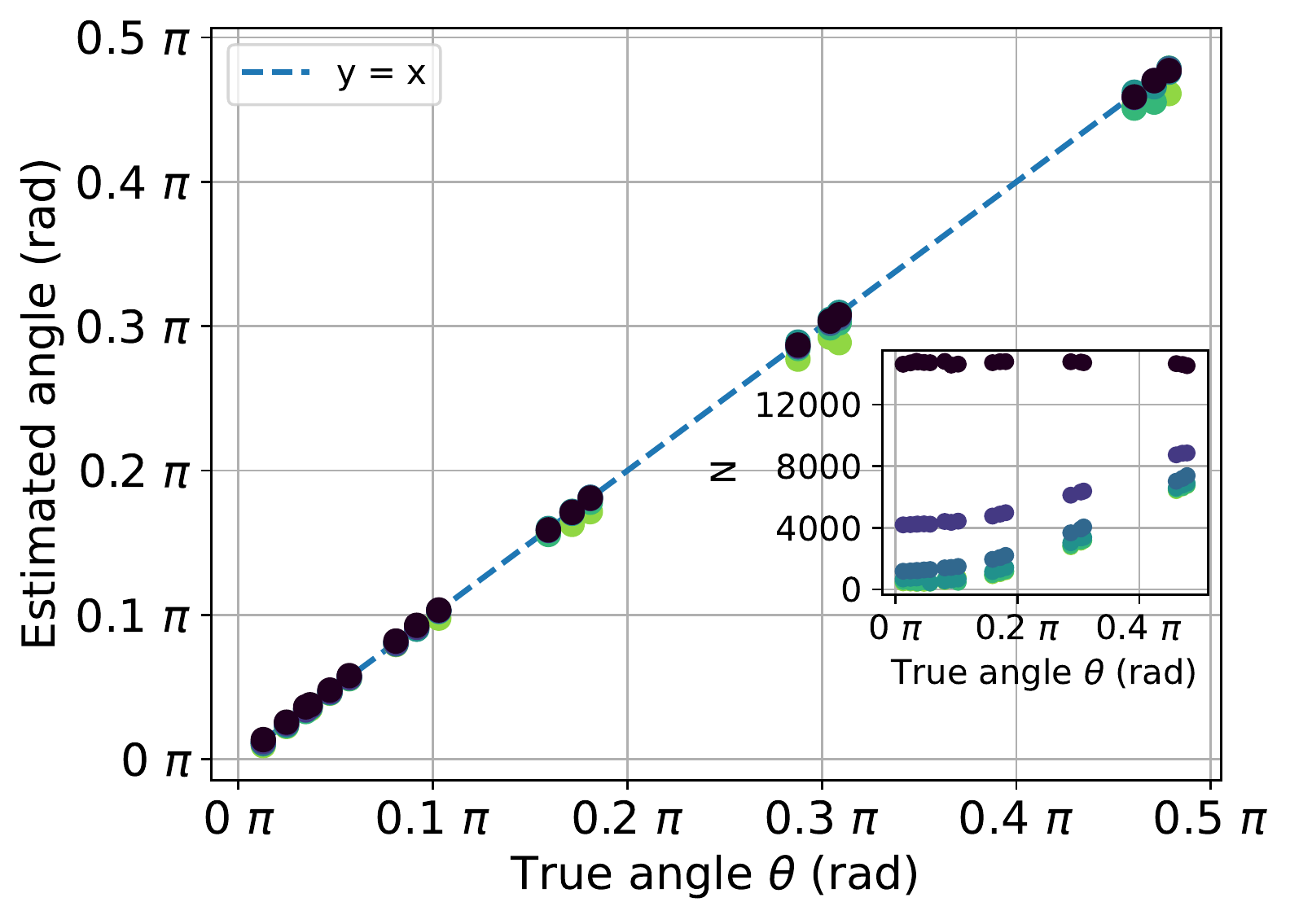}
        \caption{ }
        \label{fig_estimated_vs_true}
    \end{subfigure}
    \begin{subfigure}[b]{0.32\textwidth}
        \centering
        \includegraphics[width=\textwidth]{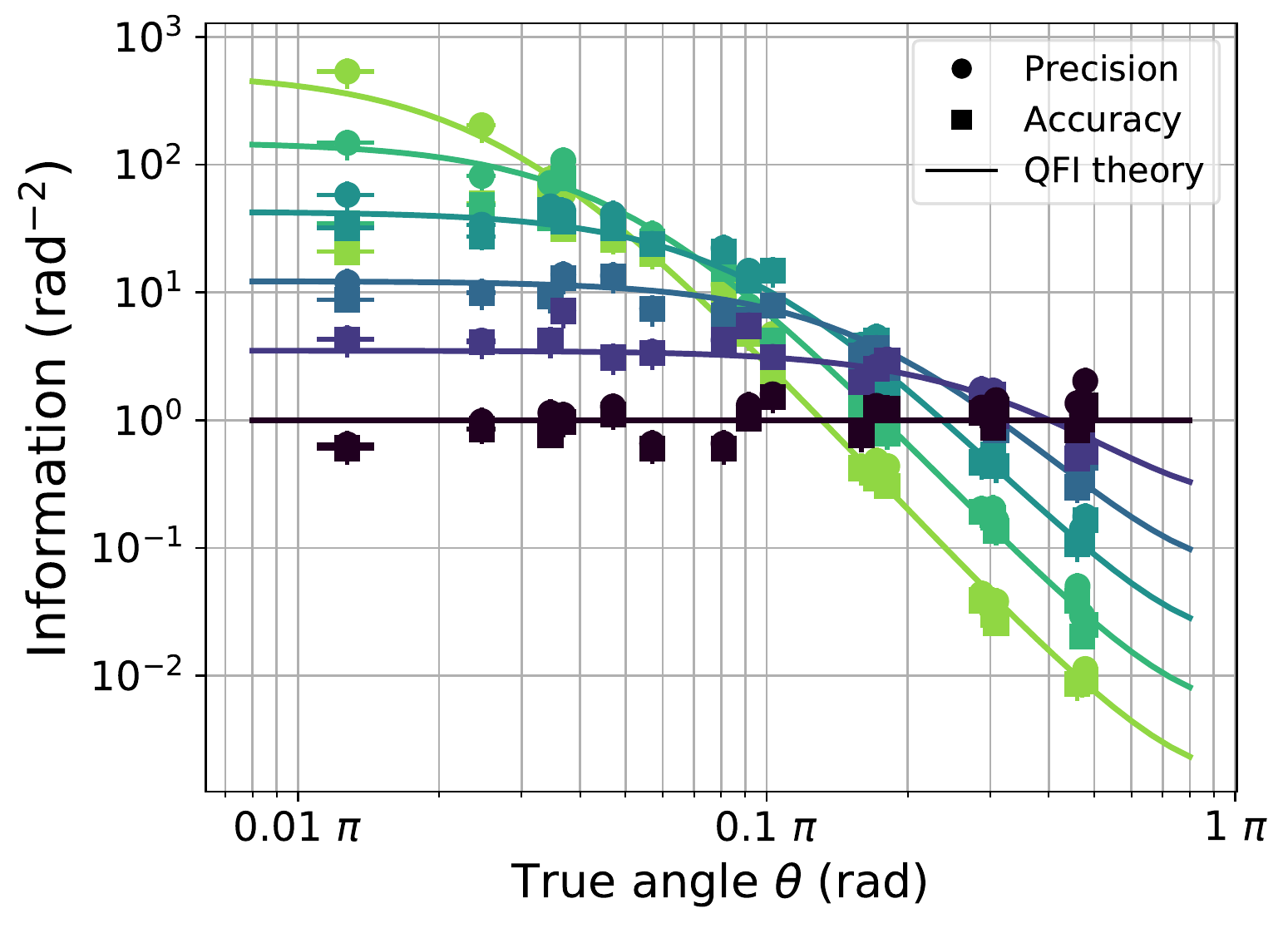}
        \caption{ }
        \label{fig_error_per_photon}
    \end{subfigure}
    \caption{Experimental performance of PPA. Color represents the postselection-parameter magnitude $|t|$.
    (a) Amplified angle vs. true angle. The slope signifies sensitivity to changes in $\theta$. The data adhere to the prediction (solid lines), $\tan(\Delta \Theta / 2) = \tan(\Delta \theta / 2) / |t|$. When $\theta$ is small 
    [$\tan(\Delta \theta / 2) \ll |t|$], 
    PPA magnifies $\theta$ by a factor $1/|t|$. 
    Setting $|t| = \tan(\Delta \theta / 2)$ amplifies $\theta$ to $\pi/2$ and optimizes the sensitivity. Decreasing $|t|$ further reduces the sensitivity, 
    rendering prior knowledge about $\theta$ important. 
    (b) Mean estimate of $\theta$ vs. true $\theta$ value. Each mean is averaged over $32$ independent estimates and agrees roughly with the true value (App. \ref{app_systematic} details systematic errors). Inset: mean detected-photon number, $N$. 
    (c) Information per photon vs. $\theta$. Solid lines represent the predicted QFI~\eqref{eq_Fisher_info_theory}. Circles represent the $32$ estimates' precision (1/variance) per mean detected photon. Squares show accuracy per detected photon (1/[mean squared error] $\equiv$ 1/MSE). All error bars follow from assuming that the variances follow a scaled $\chi^2$ distribution. The per-photon precision agrees excellently with theory and climbs to $540 \pm 150 \ \mathrm{rad}^{-2}$  at $(\theta, |t|) = (0.040 \ \mathrm{rad}, 0.044)$. The per-photon accuracy suffers from systematic errors at the smallest $\theta$ and $|t|$, yet still reaches $78 \pm 15 \ \mathrm{rad}^{-2}$ at $(\theta, |t|) = (0.116 \  \mathrm{rad}, 0.082)$.}  
\label{fig_PPA_performance}
\end{figure*}

\begin{figure*}[ht]
     %     \centering
     \begin{subfigure}[b]{0.49\textwidth}
         \centering
         \includegraphics[width=\textwidth]{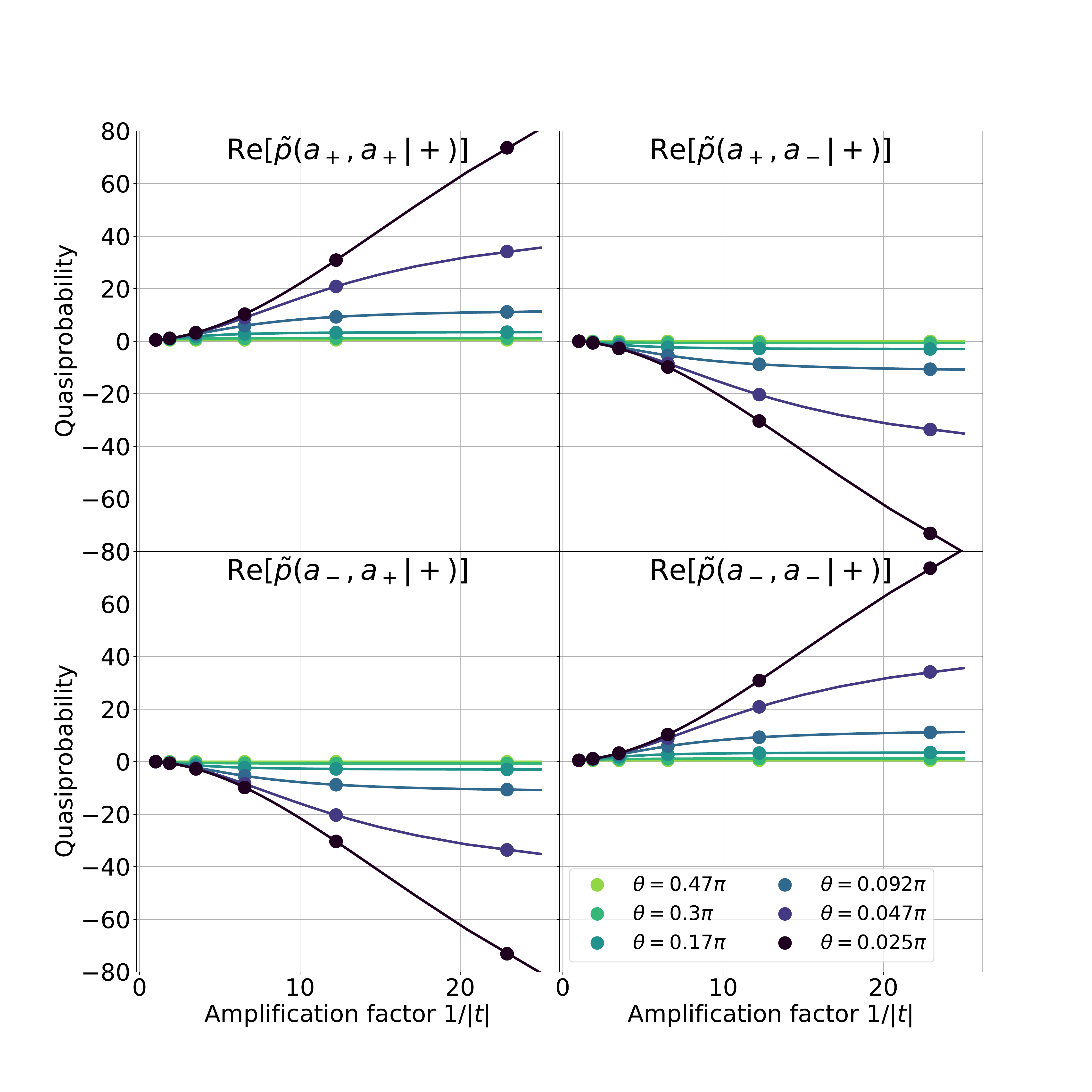}
         \caption{}
     \end{subfigure}
     \hfill
     \begin{subfigure}[b]{0.49\textwidth}
         \centering
         \includegraphics[width=\textwidth]{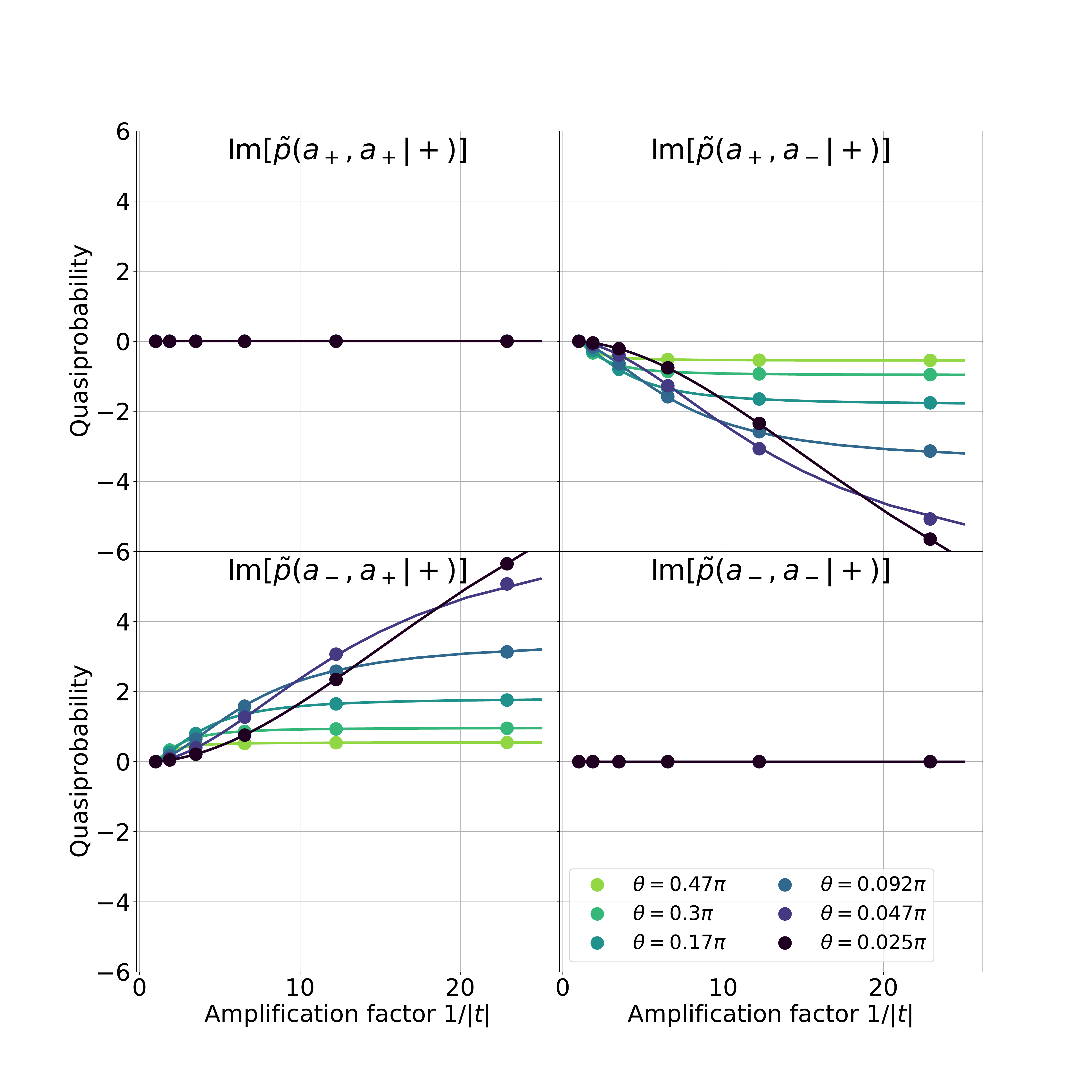}
         \caption{}
    \end{subfigure}
    \caption{Quasiprobabilities vs. amplification factor $1/|t|$. 
    We inferred the Kirkwood-Dirac distribution~\eqref{eq_Gen_Quasi},
    $\KD_{\rho(\theta), t} (a, a' | +)$, from state tomography. 
    The arguments $a_\pm = \pm 1/2$. 
    (a) Real component of distribution. (b) Imaginary component. 
    At each $\theta$ and $|t|$, the quasiprobabilities are complex, yet sum to unity. 
    As the amplification strengthens, the components' magnitudes increase. The off-diagonal quasiprobabilities' real components become extremely negative: $<-70$ at the smallest $\theta$ and $|t|$ values.}
\label{fig_KD}
\end{figure*}

\begin{figure*}[ht]
    \centering
    \begin{subfigure}[b]{0.48\textwidth}
        \centering
        \includegraphics[width=\textwidth]{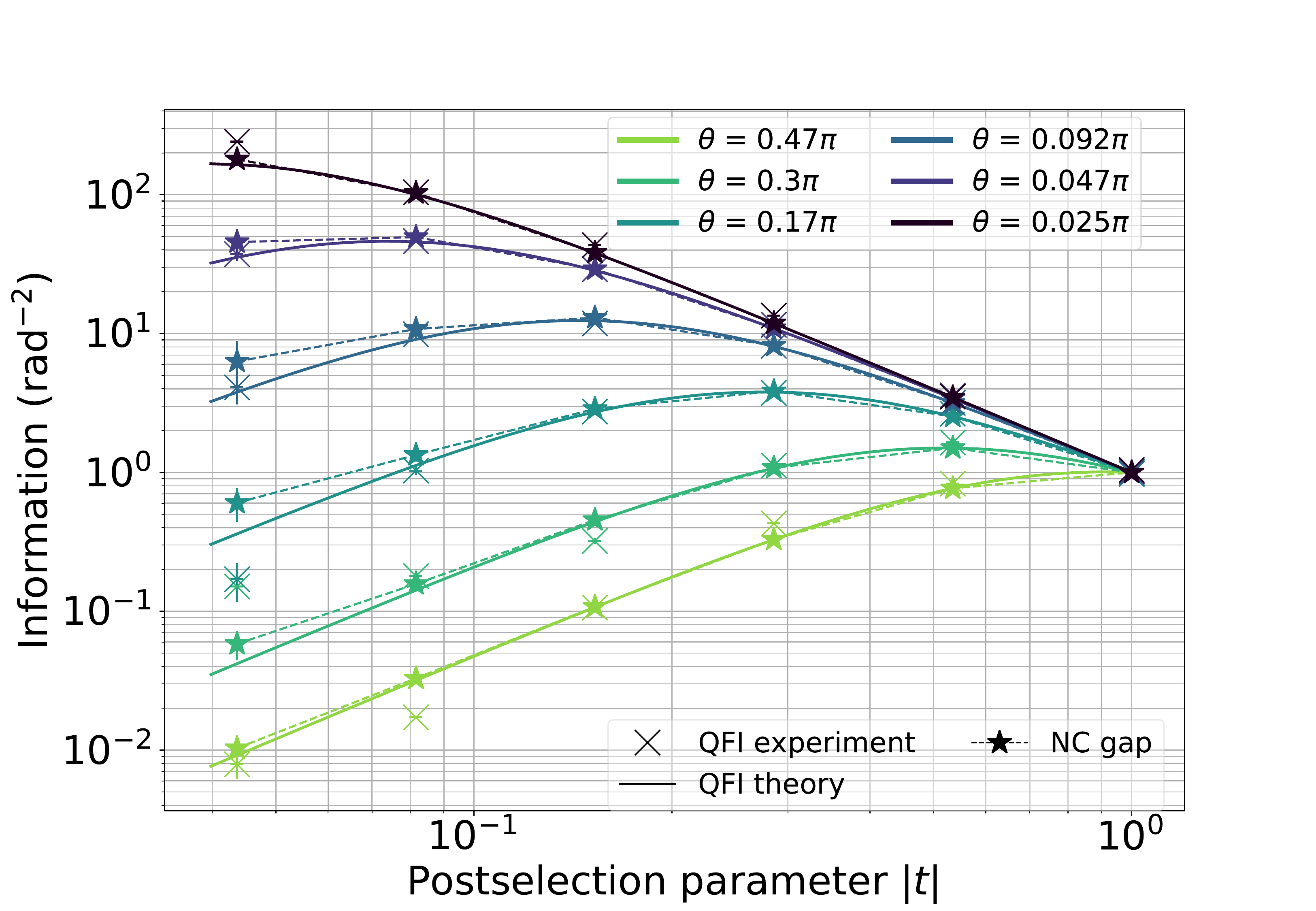}
        \caption{ }
    \end{subfigure}
    \begin{subfigure}[b]{0.48\textwidth}
        \centering
        \includegraphics[width=\textwidth]{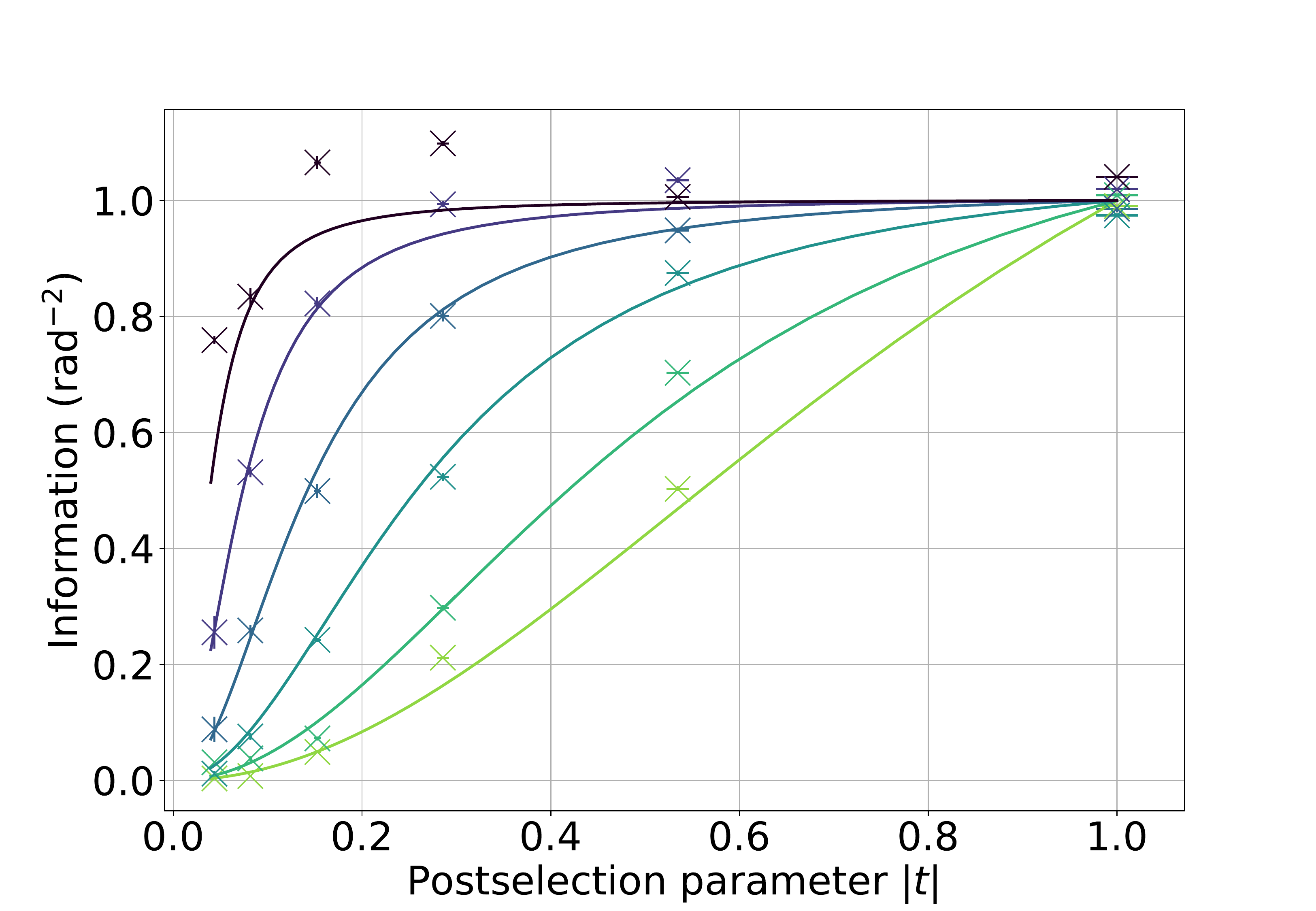}
        \caption{ }
    \end{subfigure}
    \caption{Information per detected photon (a) and information per input photon (b) vs. postselection-parameter magnitude $|t|$. Each marker shape signifies a metrological quantity. Vertical error bars denote the standard error of $4$ independent runs. In (a), each quantity is divided by the number of detected photons. In (b), each quantity is divided by the number of photons detected without postselection ($|t| = 1$).
    Crosses represent the QFI estimated from state tomography. 
    Solid lines represent the theoretical QFI prediction~\eqref{eq_Fisher_info_theory}. 
    Dashed lines with stars represent 4 times the empirical nonclassicality gap. Without postselection, our estimates are shot-noise--limited to the per--input-photon precision $1 \  \mathrm{rad}^{-2}$. As we increasingly postselect (as $|t|$ decreases), the per--detected-photon precision increases when $\theta \approx 0$ and decreases when $\tan(\theta/2) < |t|$, agreeing with theory. The smallest $|t|$ and $\theta$ provide a per--detected-photon precision 
    $> 200 \  \mathrm{rad}^{-2}$, 
    despite sacrificing little per--input-photon precision.}
\label{fig_precision_and_accuracy}
\end{figure*}

PPA involves Kraus operators, $K_+ = K(t)$ and $K_- = \sqrt{\id - K(t)^\dag K(t)}$, that effect successful and unsuccessful postselection. Table~\ref{tbl_kd_dist} shows PPA's conditional quasiprobabilities, labeled by $t$, for $\rho(\theta) = \ketbra{\Psi(\theta)}{\Psi(\theta)}$.
If $\Delta \, \theta < \pi$ and $|t|^2 < 1$,
the real part of $\KD_{\rho(\theta), t}(a_\pm, a_\mp | +)$ 
is negative,
and the postselected QFI~\eqref{eq_Fisher_info_theory} exceeds $\Delta^2$. 
This concurrence stems from an equality that we prove.

Let $x$ denote the vector of arguments for a Kirkwood-Dirac distribution $\{\KD(x)\}_x$.
Define the \emph{nonclassicality gap} %\footnote{todo: change name \nicole{Let's discuss the name before any change.}}
as the greatest difference between quasiprobabilities' absolute squares:\footnote{
% < f >
See~\cite{JRGA_19_Out,DRMAS_21_Conditions,DeBievre_21_Kirkwood} for alternative negativity measures. Equation~\eqref{eq_Prop_Result} motivates our definition.}
% < /f >
$\max_x \Set{ |\KD(x)|^2 } - \min_x \Set{ |\KD(x)|^2 }.$ 
The gap $> 1$ only if a quasiprobability $\not\in [0, 1]$.  
For any postselection operator $K_+$, 
the nonclassicality gap is proportional to 
the optimal input state's postselected QFI
(App.~\ref{app_Prove_Prop_Result}):
\begin{align}
  \label{eq_Prop_Result} &
  \QFI ( \theta )
  % % %
  = 4 \,
  \Delta^2
  \\ &  \times \nonumber
  \left[ \max_{a, a'}  \Set{  | \KD_{\rho(\theta)} (a, a' | +) |^2  }
  - \min_{a, a'}  \Set{  | \KD_{\rho(\theta)} (a, a' | +) |^2  } \right] .
\end{align}
Equation~\eqref{eq_Prop_Result} crystallizes the relationship between
postselected quantum metrology and Kirkwood-Dirac nonclassicality.

\emph{Experimental setup.---}We realize PPA in a proof-of-principle polarimetry experiment (Fig.~\ref{fig_Exp_Setup}). The to-be-estimated parameter $\theta$ is the excess birefringent phase, beyond $\pi$, imparted by a near-half--waveplate ($\textrm{HWP}0$). A heralded--single-photon source emits vertically polarized photons with wavelengths of 808 nm. The photons hit $\textrm{HWP}0$, whose optic axis lies $45^{\circ}$ above the horizontal. Tilting $\textrm{HWP}0$ through an incidence angle $\alpha$ sets its birefringent retardance to $\theta(\alpha) - \pi$. A calibration curve of $\theta(\alpha) \equiv \theta$ provides prior knowledge about $\theta$.

Denote horizontal polarization by $\ket{0}$; and vertical polarization, by $\ket{1}$.
We filter the photons by attenuating one polarization,
using an interferometer formed from polarizing beam displacers.
The postselection parameter $t$ equals (the filter's $\ket{0}$ transmission amplitude)/(the $\ket{1}$ transmission amplitude).
We control $t$ with a motorized waveplate ($\textrm{HWP}2$) placed in the interferometer.

$\textrm{HWP}0$ rotates the photon's polarization with the unitary
$\exp( i [\theta - \pi] \sigma_x /2 )$.
The generator $A = \sigma_x / 2$ has eigenvalues $a_\pm = \pm 1/2$
and eigenstates $\ket{ a_\pm } = ( \ket{0} \pm \ket{1} ) / \sqrt{2}$. The filtered photons occupy the state $\rho^\ps(\theta, t)$---ideally, the pure state~\eqref{eq_Psi_ps}. 
We projectively measure the state's polarization to estimate $\theta$.

\emph{Experimental results.---}First, we assess PPA's metrological performance. Then, we present the measured quasiprobabilities~\eqref{eq_conditional_quasi}. Comparing the quasiprobabilities with the QFI, we support Eq.~\eqref{eq_Prop_Result} experimentally.

Polarization tomography reveals how PPA boosts sensitivity. Figure~\ref{fig_amplified_vs_original} shows the postselected state's amplified angle, $\Theta$, versus the true $\theta$ value. We infer the latter using state tomography without postselection ($|t| = 1$). 
The slope of $\Theta(\theta)$
quantifies our sensitivity to small changes in $\theta$. 
When $|t| = 1$, $\Theta(\theta)$ has a unit slope. 
As we postselect more ($|t|$ decreases), the slope grows---by a factor of $> 20$ at $|t| = 0.044$.

We estimate $\theta$ by projectively measuring many copies of the amplified state identically. The measurement basis, which we optimize according to calibrations of $\theta(\alpha)$ and $t$, provides the QFI (App.~\ref{app_optimal}). 
For each $(\theta, t)$, we sample $32$ independent estimates of $\theta$. Figure~\ref{fig_estimated_vs_true} displays the mean estimate versus $\theta$. 
The mean estimates roughly agree with the true values. 
Appendix~\ref{app_systematic} details systematic errors.

Figure~\ref{fig_error_per_photon} displays our estimates' precision and accuracy, normalized by the number $N$ of detected photons. The precision per photon, $\mathrm{var}(\theta_{\rm e})^{-1} / N$, agrees excellently with the QFI~\eqref{eq_Fisher_info_theory}. The accuracy per photon, $\mathrm{MSE}(\theta_{\rm e})^{-1} / N$, mostly agrees with the QFI but falls short at the smallest $\theta$ and $|t|$. The per-photon precision enhancement maximizes at $540 \pm 150$, when $\theta = 0.040 \ \mathrm{rad}$, $|t| = 0.044$. 
The per-photon accuracy caps at $78 \pm 15$, when $\theta = 0.116 \ \mathrm{rad}$ and $|t| = 0.082$. The discrepancy between precision and accuracy arises because PPA amplifies systematic errors (App \ref{app_systematic}).

We extract the conditional quasiprobabilities~\eqref{eq_Gen_Quasi} from tomography of the unpostselected ($|t|=1$) state. Figure \ref{fig_KD} shows the quasiprobabilities' real and imaginary parts. At each $(\theta, t)$, the quasiprobabilities sum to one.
When $|t| < 1$, quasiprobabilities acquire negative real parts, so other quasiprobabilities acquire real parts $> 1$ to ensure a unit sum.
As $|t|$ decreases, elements' magnitudes increase---to $> 70$ at the smallest $\theta$ and $|t|$.

Figure~\ref{fig_precision_and_accuracy} compares the nonclassicality gap with the QFI. We compute the gap from the quasiprobabilities shown in Fig.~\ref{fig_KD}. We determine the QFI at 
$\theta = \theta_0$ empirically from estimates of 
$\rho^\ps(\theta_0, t)$ and 
$\partial \rho^\ps(\theta, t) / \partial \theta \rvert_{\theta_0}$. The derivative is the matrix slope of 
a linear fit through three tomographic estimates: 
$\rho^\ps(\theta_0 - d \theta, t), \rho^\ps(\theta_0, t),$ and $\rho^\ps(\theta_0 + d \theta, t)$; 
$d\theta = 0.035$ radians. 
We estimate the nonclassicality gap's and QFI's uncertainties by performing tomography four times and assuming Poissonian photon statistics. The empirical QFI and nonclassicality gap are consistent with the theoretical QFI [Fig.~\ref{fig_precision_and_accuracy}(a)]. Thus, our experiment corroborates the relationship~\eqref{eq_Prop_Result} between enhanced precision and quasiprobability negativity.

\emph{Conclusions.---}We have experimentally demonstrated and theoretically proved how negative Kirkwood-Dirac quasiprobabilities enhance postselected metrology.
We introduced and illustrated a scheme for phase estimation, partially postselected amplification. 
In our polarimetry experiment, PPA boosted our per--detected-photon precision by over two orders of magnitude. 
This enhancement derives from negativity of a generalized Kirkwood-Dirac quasiprobability, according to an equation that we prove and experimentally support. The negativity demonstrates that our filter provides a benefit offered by no filter that commutes with $U(\theta)$.

In theory, PPA's precision boost is unbounded. In practice, we find, the phase amplification augments systematic errors. Yet the error amplification has a silver lining, having helped us detect and correct such errors in our implementation of the generator $A$ (App.~\ref{app_systematic}).

PPA is related to weak-value amplification (WVA), a scheme for estimating couplings strengths~\cite{Vaidman88, Duck89, Hosten08, Dixon09, Pang14, Pang15, Jordan14, Starling09, Starling10, Magana14, Lyons14, Martinez17, Egan12, Harris17, Hallaji_17_Weak, Hofmann12}.
PPA and WVA concentrate information spread across many input trials into few postselected trials. 
Yet PPA differs from WVA in three ways: (i) PPA can amplify any phase, not just coupling strengths. (ii) PPA survives decoherence better. In  WVA, an interaction couples two systems. One system is measured, the other is postselected, and both must remain coherent during the interaction. PPA only requires the measured system to maintain coherence. (iii) PPA admits of a simpler mathematical treatment: WVA requires a Hilbert-space dimensionality $\geq 4$, whereas PPA works with a Hilbert-space dimensionality $\geq 2$.
PPA is therefore a promising tool for combating metrological challenges that scale with the number of completed trials. As a whole, our work interweaves the disparate studies of precision measurement and quantum foundations.

\begin{acknowledgments}
We thank Justin Dressel, Hugo Ferretti, Kent Fisher, Zhaokai Li, Seth Lloyd, and Edwin Tham for helpful conversations.  This work was supported by the Natural Sciences and Engineering Research Council (NSERC) of Canada, the National Science Foundation under Grant No. NSF PHY-1748958, the EPSRC, CIFAR, Lars Hierta's Memorial Foundation, and Girton College.   It made use of some equipment purchased through grant FQXiRFP-1819 from the Foundational Questions Institute and Fetzer Franklin Fund, a donor advised fund of Silicon Valley Community Foundation. A.M.S is a fellow of CIFAR, and N.Y.H is grateful for an NSF grant for the Institute for Theoretical Atomic, Molecular, and Optical Physics at Harvard University and the Smithsonian Astrophysical Observatory.
\end{acknowledgments}

\begin{appendices}

% To switch from two-column to one-column formatting 
\onecolumngrid

% Number subsections in the appendices as in the main text,
% except skip the capital Roman numerals.
\renewcommand{\thesection}{\Alph{section}}
\renewcommand{\thesubsection}{\Alph{section} \arabic{subsection}}
\renewcommand{\thesubsubsection}{\Alph{section} \arabic{subsection} \roman{subsubsection}}

% Label the equations in Appendix L as L1, L2, ...
\makeatletter\@addtoreset{equation}{section}
\def\theequation{\thesection\arabic{equation}}

\newpage

\section{Postselected amplification of systematic error}

\label{app_systematic}
\begin{figure}[h]
    \centering
    \includegraphics[width=0.5\textwidth]{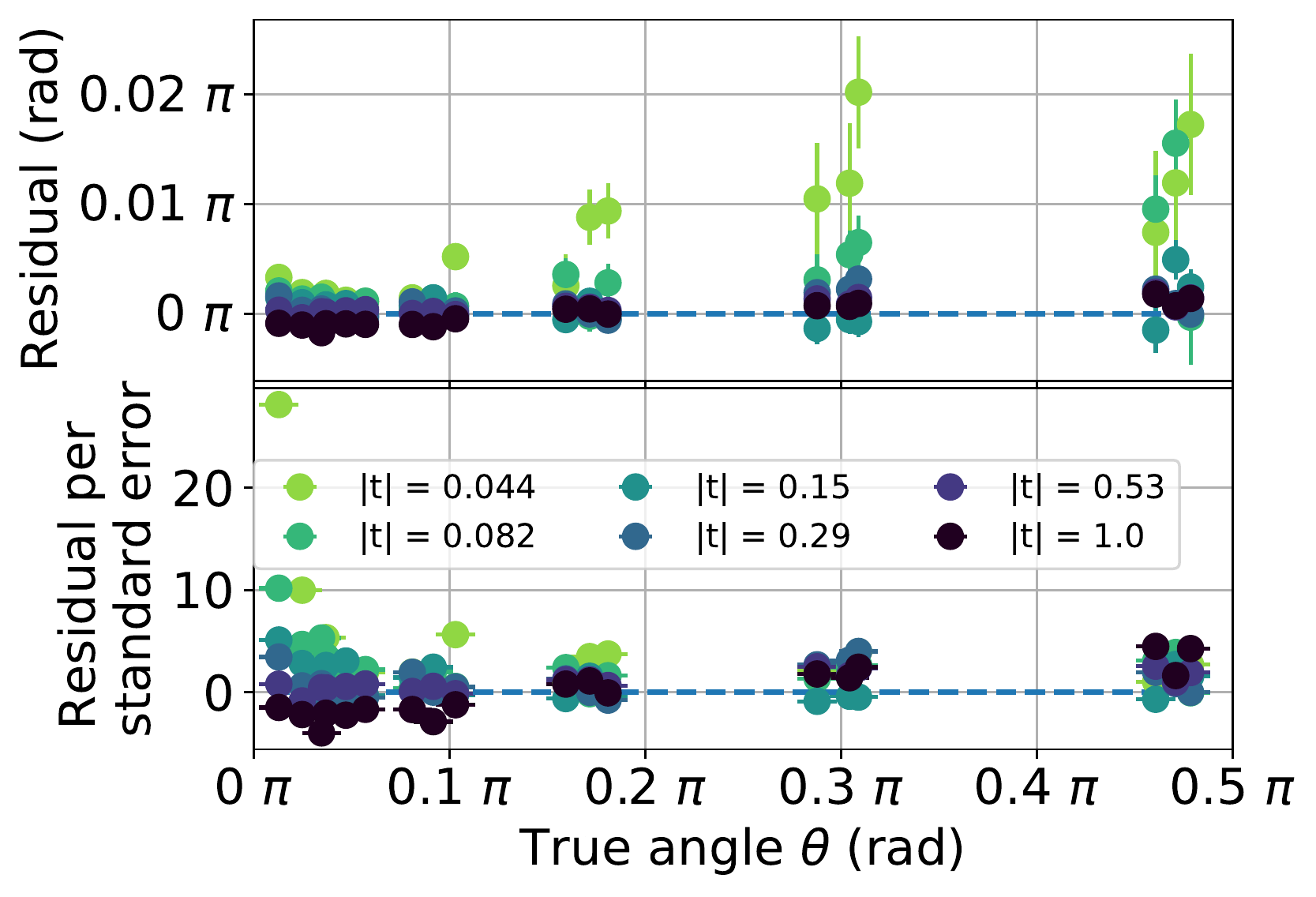}
    \caption{Difference between our estimates and the true $\theta$ value. In the lower graph, the difference is normalized by standard error.  Modest postselections ($|t| \geq 0.15$) do not suffer from significantly more systematic error than postselection-free measurements ($|t| = 1$). The two most severe postselections ($|t| = 0.044, 0.082$) amplify systematic errors significantly. }
    \label{fig_residuals}
\end{figure}

Some estimates of $\theta$ (Fig. \ref{fig_estimated_vs_true}) deviate statistically significantly from the true $\theta$ values. We now detail these deviations and explain PPA's ability to amplify certain types of systematic errors.

Figure \ref{fig_residuals} shows our mean estimates' residual, and residual per standard error, versus the true $\theta$ value. Deviations greater than $2$ standard errors point to statistically significant systematic errors. Figure \ref{fig_estimated_vs_true} showed rough agreement between our estimates and true values. However, Fig. \ref{fig_residuals} reveals that statistically significant systematic errors bedevil our estimates.

We now distinguish the systematic errors present with and without postselection. Without postselection $(|t| = 1)$, our estimates of $\theta$ differ significantly from the true (tomographically extracted) values. This error is not unique to PPA and is specific to our polarimetry experiment's details. Another type of error emerges at the two most extreme postselection levels, when $|t| \in \{0.044, 0.082\}$. The systematic errors varies strongly with $\theta$ and characterizes PPA.

While amplifying $\theta$, PPA also amplifies small errors in the implementations of 
(i) the generator $A$ and (ii) the Kraus operator $K$. 
As shown in the main text, PPA amplifies an unknown phase $\theta$ to a larger phase $\Theta$. To obtain an estimate $\theta_{\rm e}$ of $\theta$, we first obtain an estimate $\Theta_{\rm e}$  of $\Theta$. Then, we numerically invert the amplification. Inversion requires an accurate model of the experiment, but no model is perfect. In our model, postselection amplifies the state's phase to 
\begin{equation}
    \Theta = \Theta(\theta, t) 
    = 2 \arctan \LParen \tan(\theta / 2) / t \RParen,
\end{equation}
according to Eq.~\eqref{eq_Psi_ps}.
Given the amplified-angle estimate,
the $\theta$ estimate is 
$\theta_{\rm e} = \Theta^{-1}(\Theta_{\rm e}, t)$.

The setting of $t$ may suffer from a small error $\Delta t$. 
The estimated phase, as a function of the true phase, will become
\begin{equation}
    \theta_{\rm e} 
    = 2 \arctan \LParen \tan(\theta/2) 
    [1 + \Delta t / t] \RParen.
\end{equation}
The half-tangents of $\theta_{\rm e}$ and $\theta$ differ by
\begin{equation}
    \tan(\theta_{\rm e}/2) - \tan(\theta/2) 
    = [\tan(\theta/2) / t] \, \Delta t 
    = \tan(\Theta/2) \, \Delta t.
\end{equation}
This difference is proportional to the amplified half-tangent, which leads to two problems. First, amplifying $\theta$ amplifies the effects of the uncertainty in $t$. Second, the resulting systematic error cannot be corrected, because $\tan(\Theta /2)$ is unknown, by the phase-estimation task's nature.

Separately, errors can mar the implementation of the generator $A$. Ideally, $A = \sigma_x/2$. However, $A$ is effected by a waveplate whose optic axis may be
% if the orientation of the optic axis of the waveplate that implements $A$ is 
rotated by a small angle $\epsilon$ from $45^{\circ}$ to the horizontal. The generator will become 
$A(\epsilon) = \cos(2 \epsilon) \sigma_x/2 + \sin(2 \epsilon) \sigma_z/2$. 
Under $e^{i (\theta - \pi) A(\epsilon)}$, the initial state $\ket{1}$ evolves to
\begin{equation}
    \ket{\psi(\epsilon)} 
    = \cos(\theta/2) [\cos(2 \epsilon) \ket{0} - \sin(2 \epsilon) \ket{1}] +i \sin(\theta/2) \ket{1}.
\end{equation}
This state's polar half-angle has a tangent
% The tangent of this state's polar half-angle is
\begin{equation}
    \frac{|\braket{1}{\psi(\epsilon)}|}{|\braket{0}{\psi(\epsilon)}|} = \sqrt{\frac{\sin^2(2\epsilon) + \tan^2(\theta/2)}{\cos^2(2\epsilon)}} \, .
\end{equation}
When $\sin^2(2\epsilon) \ll \tan^2(\theta/2)$, 
$\epsilon$ affects the polar angle negligibly.
However, if $\sin^2(2\epsilon) > \tan^2(\theta/2)$, 
the polar angle depends on $\epsilon$ considerably. 
Amplifying the polar angle with PPA amplifies also
the error in the $A$ implementation.

One can compensate, or correct, for the error amplification. Moreover, we exploited it to improve our $A$ implementation while aligning our equipment. However, calibration and correction errors propagate to $\theta$ eventually. Thus, the precision of one's $K(t)$ and $A$ implementations limits how much one can filter and postselect usefully. This phenomenon extends beyond our polarimetery experiment to every PPA implementation.

\section{Optimal measurement}
\label{app_optimal}

We estimated $\theta$ with an optimal polarization measurement. By \emph{optimal measurement}, we mean a measurement whose information yield, averaged over many trials, equals the QFI. 
Let $\Lambda$ denote the symmetric logarithmic derivative of $\rho$, defined implicitly through the Sylvester equation,
\begin{equation}
\label{eq_SLD}
    \frac{\partial \rho}{\partial \theta} = \frac{\Lambda \rho + \rho \Lambda}{2}.
\end{equation}
Projectively measuring an eigenbasis of $\Lambda$ is optimal~\cite{Braunstein94}.
Therefore, we aim to solve Eq.~\eqref{eq_SLD} for $\Lambda$.

We obtain a solution as follows.
Let $\mathrm{vec}(Y)$ denote the vectorized form 
of a matrix $Y$ in row-major order. 
For matrices $X$, $Y$, and $Z$ with appropriate dimensions,
$\mathrm{vec}(X Y Z) 
= (X \otimes Z^T) \mathrm{vec}(Y)$. 
Let $Y^{+}$ denote the Moore-Penrose inverse of a matrix $Y$.
Solving for $\Lambda$, we vectorize it:
\begin{equation}
    \mathrm{vec}(\Lambda) = \left ( \frac{\rho \otimes \id + \id \otimes \rho^T}{2} \right )^{+} \mathrm{vec} \left (\frac{\partial \rho}{\partial \theta} \right ).
\end{equation}

In our experiment, the photon polarization is nearly, but not exactly, pure:
The initial polarization is 
$\rho_0 = v \ketbra{0}{0} + (1-v) \id/2$.
The visibility $v$ is close, but not equal, to unity. 
The postselected state is
\begin{equation}
    \rho^\ps(\theta, t) = K(t) U(\theta) \rho_0 U(\theta)^\dagger K(t)^\dagger / p^\ps(\theta, t),
\end{equation}
and the postselection probability is
\begin{equation}
    p^\ps(\theta, t) = \Tr \LParen 
    K(t) U(\theta) \rho_0 U(\theta)^\dagger K(t)^\dagger \RParen.
\end{equation}
$\Lambda$ assumes the form
\begin{equation}
    \Lambda = v \, p^\ps(\theta, t) 
    \left[ \frac{1 - |t|^2}{2} \sin(\theta) \id + \cos(\theta) (Re[t] \sigma_x + Im[t] \sigma_y) + \frac{1 + |t|^2}{2} \sin(\theta) \sigma_z  \right].
\end{equation}
An optimal measurement is thus a polarization onto the Bloch vector whose polar angle $\theta_\mathrm{opt}$ and azimuthal angle $\phi_\mathrm{opt}$ are defined through
\begin{equation}
    \cot(\theta_\mathrm{opt}) = \frac{1 + |t|^2}{2 |t|} \tan(\theta), \ \ \ \ \tan(\phi_\mathrm{opt}) = \Im[t] / \Re[t].
\end{equation}
$\theta_\mathrm{opt}$ turns out to be independent of $v$ but not of $\theta$.
We use prior knowledge about $\theta$, derived from our calibration curve $\theta(\alpha)$, to estimate $\theta_{\mathrm{opt}}$ and approximate an optimal measurement.

\section{Generalized Kirkwood-Dirac distribution}
\label{app_KD}

We called the quasiprobabilities~\eqref{eq_Gen_Quasi} ``Kirkwood-Dirac quasiprobabilities.'' 
This appendix justifies that terminology.
Kirkwood and Dirac independently defined a quasiprobability
dependent on a pure state $\ket{\psi}$ 
and on two projectors, $\ketbra{a}{a}$ and $\ketbra{f}{f}$~\cite{Kirkwood33,Dirac45}:
$\Tr ( | f \rangle \langle f | a \rangle \langle a | 
\psi \rangle \langle \psi |)$.
Wiseman generalized the pure state to 
an arbitrary quantum state $\rho$~\cite{Wiseman_02_Weak}: 
$\Tr ( | f \rangle \langle f | a \rangle \langle a | \rho)$.
Yunger Halpern and collaborators generalized to
arbitrarily many projectors $\Pi^{M_j}_{m_j}$,
which can project onto multidimensional subspaces~\cite{NYH_17_Jaryznski, NYH_18_Quasiprobability}:
$\Tr \left( \Pi^{(k)}_{m_k} \Pi^{(k-1)}_{m_{k-1}} \ldots \Pi^{(1)}_{m_1} \rho \right)$.
They called this quantity an \emph{extended Kirkwood-Dirac quasiprobability}.

We generalize the projectors to elements of POVMs
$\Set{ M^\1_{m_1} }_{m_1}, \Set{ M^\2_{m_2} }_{m_2}, \ldots, 
\Set{ M^{(k)}_{m_k} }_{m_k}$:
\begin{align}
   \tilde{p}_\rho (m_1, m_2, \ldots, m_k)
   := \Tr \left( M^{(k)}_{m_k}  M^{(k-1)}_{m_{k-1} } \ldots M^\1_{m_1} \rho \right) .
   \label{eq_KD_multi_arg}
\end{align}
By definition, the $\ell^\th$ POVM's elements are positive-semidefinite operators that resolve unity: $M^{(\ell)}_{m_\ell} \geq 0$, and $\sum_{m_\ell} M_{m_\ell}=\id$. 
We could call our quasiprobability~\eqref{eq_KD_multi_arg} a ``generalized extended Kirkwood-Dirac quasiprobability''; but the name, itself, would be too extended. 
Following Wiseman's lead, we avoid new names; 
we christen our construction, and rechristen the extended Kirkwood-Dirac quasiprobability, a ``Kirkwood-Dirac quasiprobability.''

To merit the name, our quasiprobability distribution should exhibit two recursive properties:
\begin{enumerate}
   \item \label{item_Prob}
   A Kirkwood-Dirac distribution with one argument is a probability distribution (albeit a trivial one).
   
   \item \label{item_Marg}
   Marginalizing over any argument of a Kirkwood-Dirac distribution with $k > 1$ arguments yields a Kirkwood-Dirac distribution with $k - 1$ arguments.
\end{enumerate}
Consequently, marginalizing over $k - 1$ arguments of a $k$-argument Kirkwood-Dirac distribution yields a probability distribution. We now verify that our distribution~\eqref{eq_KD_multi_arg} has these properties.

Consider the one-argument distribution with elements 
$\KD_\rho(m_1) = \Tr(M^\1_{m_1} \rho)$. It satisfies property~\ref{item_Prob} because (i) each $M^\1_{m_1}$ is positive-semidefinite, so $\KD_\rho (m_1) \geq 0$; and (ii) $\Set{ M^\1_{m_1} }_{m_1}$ resolves unity, so $\sum_{m_1} \KD_\rho (m_1) = \Tr(\rho) = 1$. The element $\KD_\rho(m_1)$ is the probability that, 
upon preparing $\rho$ and measuring $\Set{ M^\1_{m_1} }_{m_1}$,
one obtains outcome $m_1$.

Now, consider the $k$-argument distribution~\eqref{eq_KD_multi_arg}. Marginalizing over the $\ell^\th$ POVM yields
\begin{align}
    \sum_{m_\ell} \KD_\rho (m_1, m_2, \ldots, m_{\ell - 1}, m_{\ell}, m_{\ell + 1}, \ldots, m_k)
    & = \KD_\rho(m_1, m_2, \ldots m_{\ell-1}, m_{\ell+1}, m_{\ell+2}, \ldots, m_k) \\ 
    & = \Tr \left( M^{(k)}_{m_k} 
    M^{(k-1)}_{m_{k-1}} \ldots 
    M^{(\ell+1)}_{m_{\ell+1} } 
    M^{(\ell-1)}_{m_{\ell-1} } 
    M^{(\ell-2)}_{m_{\ell-2} } \ldots   
    M^\1_{m_1} \rho \right),
\end{align}
because each POVM resolves unity. The right-hand side is a $(k-1)$-argument Kirkwood-Dirac distribution. Therefore, our construction has property~\ref{item_Marg}. 
% NYH: Repetitive. The text below has already been stated above, below property 2.
% Consequently, marginalizing $\KD_\rho (m_1, m_2, \ldots, m_k)$ over all POVMs but the $\ell^\th$ produces the distribution with elements $\KD_\rho (m_\ell) = \Tr(M^{(\ell)}_{m_\ell} \rho)$.

%
%
%
\section{Proof of proportionality between postselected quantum Fisher information and quasiprobability nonclassicality}
\label{app_Prove_Prop_Result}

% Reference 1: DRMAS proved the claim for nondegenerate observables A on June 22, 2020.
% Reference 2: NYH generalized David's argument to degenerate observables A on June 23, 2020.

Equation~\eqref{eq_Prop_Result} interrelates 
the postselected quantum Fisher information
and the generalized quasiprobability's nonclassicality.
We prove this equation---in fact, a generalization of the equation---here.

\textbf{Generalized setup:}
All the definitions presented here are independent of
the definitions presented in the main text.
For example, the $A$ defined here is independent of
the $A$ defined in the main text.
However, quantities defined in the main text are examples of quantities defined here.

During a parameter-estimation experiment, 
the input state $\rho$ undergoes a unitary
$U(\theta) = e^{i \theta A}$ generated by a Hermitian operator $A$:
\begin{align}
  \label{eq_Rho_Evolve}
  \rho  \mapsto  U(\theta) \rho U(\theta)^\dag  =  \rho(\theta) .
\end{align}
The parameter to be estimated is $\theta$. The system corresponds to a Hilbert space of arbitrary dimensionality.
We assume that $\rho$ is a pure state with support on only two $A$ eigenspaces. The projectors onto these eigenspaces, we denote by $\Pi_\Min$ and $\Pi_\Max$. Their associates eigenvalues are $a_\Min$ and $a_\Max > a_\Min$.

After undergoing the unitary, the state meets a filter. If the state survives the filter, a Kraus operator $K_+$
updates the state. 
Otherwise, $K_- = \sqrt{\id - K_+^\dag K_+}$ updates the state.
The state survives the filter with the postselection probability 
$p^\ps(\theta)  
=  \Tr \LParen \rho (\theta) \, K_+^\dag K_+ \RParen$.
We can represent the experiment with the Kirkwood-Dirac quasiprobability
\begin{align}
  \KD_{\rho(\theta)}  (a, f, a')
  =  \Tr \LParen  \Pi_{a'}  \,  K_f^\dag K_f  \,  \Pi_a  \,  \rho(\theta) \RParen ,
  % % %
  \quad \text{for }
  a, a' \in \{ a_\Max, a_\Min \} .
\end{align}
Our result depends on the conditional quasiprobabilities, 
labeled by $f = +$,
associated with a successful postselection:
\begin{equation}
    \KD_{\rho(\theta)} (a, a' | +) = \KD_{\rho(\theta)} (a, +, a') / p^\ps(\theta) .
\end{equation}

We assume that two ``diagonal'' quasiprobabilities, labeled by $a = a'$, equal each other:
\begin{align}
  \label{eq_2_Quasiprobs_Equal}
  \Tr \LParen \Pi_\Max  \,  \rho(\theta)  \,  
  \Pi_\Max  \,  K_+^\dag K_+ \RParen
  =  \Tr  \LParen  \Pi_\Min  \,  \rho(\theta)  \,  
  \Pi_\Min  \,  K_+^\dag K_+ \RParen .
\end{align}
This condition is satisfied by the main text's qubit system of interest, $A = \sigma_x / 2$, and input state 
$( \ket{a_+}  +  \ket{a_-} ) / 2  =  \ket{0}$.

\textbf{Generalized equality:}
The postselected QFI, $\QFI (\theta)$, is proportional to 
the quasiprobability's nonclassicality gap:
\begin{align}
  \label{eq_Prop_Result_Gen}
  \QFI ( \theta)
  % % %
  =  4  \,
  (a_\Max  -  a_\Min )^2
  \left(  \max_{a, a'}  \Set{ |  \KD_{\rho(\theta)} (a, a' | +)  |^2  }
  -  \min_{a, a'}  \Set{ | \KD_{\rho(\theta)}  (a, a' | +)  |^2 }
  \right) .
\end{align}

\textbf{Proof:}
We calculate postselected quantum Fisher information 
by substituting from Eq.~\eqref{eq_Rho_Evolve} into a formula for a pure state's postselected QFI \{Eq.~(5) in~\cite{DRMAS_20_Quantum}\}:
\begin{align}
  \QFI ( \theta )
  % % %
  & = \frac{4}{ p^\ps(\theta) }  \,
  \Tr \LParen A  \,  \rho(\theta)  \,  A \,  K_+^\dag K_+ \RParen
  -  \frac{4}{ [ p^\ps(\theta) ]^2 }  \,
  \lvert  \Tr \LParen A  \,  \rho(\theta)  \,  K_+^\dag K_+ \RParen \rvert^2  \\
  % % %
  \label{eq_Prop_App_Help1}
  & = \frac{4}{ [ p^\ps(\theta) ]^2 }  \,  \left[
  \Tr \LParen A  \,  \rho(\theta)  \,  A \,  K_+^\dag K_+ \RParen  \,
  \Tr \LParen \rho(\theta)  \,  K_+^\dag K_+ \RParen
  -  \Tr \LParen A  \,  \rho(\theta)  \,  K_+^\dag K_+ \RParen  \,
  \Tr \LParen \rho(\theta)  \,  A \,  K_+^\dag K_+ \RParen
  \right] .
\end{align}
We will rewrite the right-hand side, by decomposing the evolved state as
\begin{align}
  \rho(\theta)
  % % %
  =  ( \Pi_\Max  +  \Pi_\Min )  \,  \rho(\theta)  \,
  ( \Pi_\Max  +  \Pi_\Min ) \\
  % % %
  & =  \Pi_\Max \,  \rho(\theta) \,  \Pi_\Max
  +  \Pi_\Max \,  \rho(\theta) \,  \Pi_\Min
  +  \Pi_\Min \,  \rho(\theta) \,  \Pi_\Max
  +  \Pi_\Min \,  \rho(\theta) \,  \Pi_\Min .
\end{align}
We substitute this expression into Eq.~\eqref{eq_Prop_App_Help1}.
Then, we invoke the eigenvalue equations
$A \Pi_\Max  =  a_\Max  \Pi_\Max$ and
$A \Pi_\Min  =  a_\Min  \Pi_\Min$.
Once we multiply out, terms cancel:
\begin{align}
  \QFI ( \theta  |  \rho(\theta) )
  % % %
  & = \frac{4}{ [ p^\ps(\theta) ]^2 }  \,
  ( a_\Max  -  a_\Min )^2
  \big[  \Tr \LParen \Pi_\Max  \,  \rho(\theta)  \,  
  \Pi_\Max  \,  K_+^\dag K_+ \RParen  \,
  \Tr \LParen \Pi_\Min  \,  \rho(\theta)  \,  
  \Pi_\Min \,  K_+^\dag K_+ \RParen
  \nonumber \\ & \qquad \qquad \qquad \qquad \qquad \quad
  -  \Tr \LParen \Pi_\Max  \,  \rho(\theta)  \,  
              \Pi_\Min \,  K_+^\dag K_+ \RParen  \,
  \Tr \LParen \Pi_\Min \,  \rho(\theta)  \,  
          \Pi_\Max \,  K_+^\dag K_+ \RParen
  \big]  .
\end{align}
The final term equals 
$\lvert \Tr \LParen \Pi_\Max  \,  \rho(\theta) \,  
\Pi_\Min \,  K_+^\dag K_+ \RParen \rvert^2$, so
\begin{align}
  \label{eq_Prop_App_Help2}
  \QFI ( \theta )
  % % %
  & =  4  \,
  ( a_\Max  -  a_\Min )^2  
  \left[  \KD_{ \rho(\theta) }  ( a_\Max,  a_\Max | + )  \,
  \KD_{ \rho(\theta) }  ( a_\Min ,  a_\Min | + )
  - |  \KD_{ \rho(\theta) }  ( a_\Max,  a_\Min | + )  |^2
  \right]  .
\end{align}

Since $\QFI ( \theta )$ and 
$(a_\Max  -  a_\Min)^2  \geq 0$, 
Eq.~\eqref{eq_Prop_App_Help2} implies
\begin{align}
  \label{eq_Prop_App_Help3}
  \KD_{ \rho(\theta) }  ( a_\Max,  a_\Max | + )  \,
  \KD_{ \rho(\theta) }  ( a_\Min ,  a_\Min | +  )
  % % %
  \geq  |  \KD_{ \rho(\theta) }  ( a_\Max,  a_\Min | + )  |^2 .
\end{align}
The left-hand side equals
$\left[ \KD_{ \rho(\theta) }  (a_\Max,  a_\Max | +)  \right]^2$,
by Eq.~\eqref{eq_2_Quasiprobs_Equal}.
Since each side of Eq.~\eqref{eq_2_Quasiprobs_Equal} is overtly a probability,
$\KD_{ \rho(\theta) }  (a_\Max,  a_\Max | +)$ is real.
The square therefore equals the square modulus.
Substituting into Ineq.~\eqref{eq_Prop_App_Help3} yields
$|  \KD_{\rho(\theta)}  (a_\Max,  a_\Max | +)  |^2
\geq  |  \KD_{\rho(\theta) }  (a_\Max,  a_\Min | +)  |^2$.
This inequality contains the square moduli of
all the quasiprobabilities for which $f = +$.
Hence the inequality's left-hand side equals the greatest square probability,
$\max_{a, a'}  \Set{
[  \KD_{\rho(\theta)}  (a,  a' | +)  ]^2  }$,
while the right-hand side equals the least square modulus,
$\min_{a, a'}  \Set{
[  \KD_{\rho(\theta)}  (a,  a' | +)  ]^2  }$.
Substituting the max and min into Eq.~\eqref{eq_Prop_App_Help2}
yields Eq.~\eqref{eq_Prop_Result_Gen}. $\square$

\end{appendices}

\bibliography{refs}

\begin{thebibliography}{78}%
\makeatletter
\providecommand \@ifxundefined [1]{%
 \@ifx{#1\undefined}
}%
\providecommand \@ifnum [1]{%
 \ifnum #1\expandafter \@firstoftwo
 \else \expandafter \@secondoftwo
 \fi
}%
\providecommand \@ifx [1]{%
 \ifx #1\expandafter \@firstoftwo
 \else \expandafter \@secondoftwo
 \fi
}%
\providecommand \natexlab [1]{#1}%
\providecommand \enquote  [1]{``#1''}%
\providecommand \bibnamefont  [1]{#1}%
\providecommand \bibfnamefont [1]{#1}%
\providecommand \citenamefont [1]{#1}%
\providecommand \href@noop [0]{\@secondoftwo}%
\providecommand \href [0]{\begingroup \@sanitize@url \@href}%
\providecommand \@href[1]{\@@startlink{#1}\@@href}%
\providecommand \@@href[1]{\endgroup#1\@@endlink}%
\providecommand \@sanitize@url [0]{\catcode `\\12\catcode `\$12\catcode
  `\&12\catcode `\#12\catcode `\^12\catcode `\_12\catcode `\%12\relax}%
\providecommand \@@startlink[1]{}%
\providecommand \@@endlink[0]{}%
\providecommand \url  [0]{\begingroup\@sanitize@url \@url }%
\providecommand \@url [1]{\endgroup\@href {#1}{\urlprefix }}%
\providecommand \urlprefix  [0]{URL }%
\providecommand \Eprint [0]{\href }%
\providecommand \doibase [0]{https://doi.org/}%
\providecommand \selectlanguage [0]{\@gobble}%
\providecommand \bibinfo  [0]{\@secondoftwo}%
\providecommand \bibfield  [0]{\@secondoftwo}%
\providecommand \translation [1]{[#1]}%
\providecommand \BibitemOpen [0]{}%
\providecommand \bibitemStop [0]{}%
\providecommand \bibitemNoStop [0]{.\EOS\space}%
\providecommand \EOS [0]{\spacefactor3000\relax}%
\providecommand \BibitemShut  [1]{\csname bibitem#1\endcsname}%
\let\auto@bib@innerbib\@empty
%</preamble>
\bibitem [{\citenamefont {Rozema}\ \emph {et~al.}(2012)\citenamefont {Rozema},
  \citenamefont {Darabi}, \citenamefont {Mahler}, \citenamefont {Hayat},
  \citenamefont {Soudagar},\ and\ \citenamefont {Steinberg}}]{Rozema12-2}%
  \BibitemOpen
  \bibfield  {author} {\bibinfo {author} {\bibfnamefont {L.~A.}\ \bibnamefont
  {Rozema}}, \bibinfo {author} {\bibfnamefont {A.}~\bibnamefont {Darabi}},
  \bibinfo {author} {\bibfnamefont {D.~H.}\ \bibnamefont {Mahler}}, \bibinfo
  {author} {\bibfnamefont {A.}~\bibnamefont {Hayat}}, \bibinfo {author}
  {\bibfnamefont {Y.}~\bibnamefont {Soudagar}},\ and\ \bibinfo {author}
  {\bibfnamefont {A.~M.}\ \bibnamefont {Steinberg}},\ }\bibfield  {title}
  {\bibinfo {title} {Violation of heisenberg's measurement-disturbance
  relationship by weak measurements},\ }\href
  {https://doi.org/10.1103/PhysRevLett.109.100404} {\bibfield  {journal}
  {\bibinfo  {journal} {Phys. Rev. Lett.}\ }\textbf {\bibinfo {volume} {109}},\
  \bibinfo {pages} {100404} (\bibinfo {year} {2012})}\BibitemShut {NoStop}%
\bibitem [{\citenamefont {Giovannetti}\ \emph {et~al.}(2004)\citenamefont
  {Giovannetti}, \citenamefont {Lloyd},\ and\ \citenamefont
  {Maccone}}]{giovannetti2004quantum}%
  \BibitemOpen
  \bibfield  {author} {\bibinfo {author} {\bibfnamefont {V.}~\bibnamefont
  {Giovannetti}}, \bibinfo {author} {\bibfnamefont {S.}~\bibnamefont {Lloyd}},\
  and\ \bibinfo {author} {\bibfnamefont {L.}~\bibnamefont {Maccone}},\
  }\bibfield  {title} {\bibinfo {title} {Quantum-enhanced measurements: beating
  the standard quantum limit},\ }\href@noop {} {\bibfield  {journal} {\bibinfo
  {journal} {Science}\ }\textbf {\bibinfo {volume} {306}},\ \bibinfo {pages}
  {1330} (\bibinfo {year} {2004})}\BibitemShut {NoStop}%
\bibitem [{\citenamefont {Giovannetti}\ \emph {et~al.}(2006)\citenamefont
  {Giovannetti}, \citenamefont {Lloyd},\ and\ \citenamefont
  {Maccone}}]{Giovanetti06}%
  \BibitemOpen
  \bibfield  {author} {\bibinfo {author} {\bibfnamefont {V.}~\bibnamefont
  {Giovannetti}}, \bibinfo {author} {\bibfnamefont {S.}~\bibnamefont {Lloyd}},\
  and\ \bibinfo {author} {\bibfnamefont {L.}~\bibnamefont {Maccone}},\
  }\bibfield  {title} {\bibinfo {title} {Quantum metrology},\ }\href@noop {}
  {\bibfield  {journal} {\bibinfo  {journal} {Physical review letters}\
  }\textbf {\bibinfo {volume} {96}},\ \bibinfo {pages} {010401} (\bibinfo
  {year} {2006})}\BibitemShut {NoStop}%
\bibitem [{\citenamefont {Giovannetti}\ \emph {et~al.}(2011)\citenamefont
  {Giovannetti}, \citenamefont {Lloyd},\ and\ \citenamefont
  {Maccone}}]{Giovanetti11}%
  \BibitemOpen
  \bibfield  {author} {\bibinfo {author} {\bibfnamefont {V.}~\bibnamefont
  {Giovannetti}}, \bibinfo {author} {\bibfnamefont {S.}~\bibnamefont {Lloyd}},\
  and\ \bibinfo {author} {\bibfnamefont {L.}~\bibnamefont {Maccone}},\
  }\bibfield  {title} {\bibinfo {title} {Advances in quantum metrology},\
  }\href@noop {} {\bibfield  {journal} {\bibinfo  {journal} {Nature photonics}\
  }\textbf {\bibinfo {volume} {5}},\ \bibinfo {pages} {222} (\bibinfo {year}
  {2011})}\BibitemShut {NoStop}%
\bibitem [{\citenamefont {Polino}\ \emph {et~al.}(2020)\citenamefont {Polino},
  \citenamefont {Valeri}, \citenamefont {Spagnolo},\ and\ \citenamefont
  {Sciarrino}}]{Polino20}%
  \BibitemOpen
  \bibfield  {author} {\bibinfo {author} {\bibfnamefont {E.}~\bibnamefont
  {Polino}}, \bibinfo {author} {\bibfnamefont {M.}~\bibnamefont {Valeri}},
  \bibinfo {author} {\bibfnamefont {N.}~\bibnamefont {Spagnolo}},\ and\
  \bibinfo {author} {\bibfnamefont {F.}~\bibnamefont {Sciarrino}},\ }\bibfield
  {title} {\bibinfo {title} {Photonic quantum metrology},\ }\href
  {https://doi.org/10.1116/5.0007577} {\bibfield  {journal} {\bibinfo
  {journal} {AVS Quantum Science}\ }\textbf {\bibinfo {volume} {2}},\ \bibinfo
  {pages} {024703} (\bibinfo {year} {2020})},\ \Eprint
  {https://arxiv.org/abs/https://doi.org/10.1116/5.0007577}
  {https://doi.org/10.1116/5.0007577} \BibitemShut {NoStop}%
\bibitem [{\citenamefont {Yoon}\ \emph {et~al.}(2020)\citenamefont {Yoon},
  \citenamefont {Lee}, \citenamefont {Rockstuhl}, \citenamefont {Lee},\ and\
  \citenamefont {Lee}}]{yoon2020experimental}%
  \BibitemOpen
  \bibfield  {author} {\bibinfo {author} {\bibfnamefont {S.-J.}\ \bibnamefont
  {Yoon}}, \bibinfo {author} {\bibfnamefont {J.-S.}\ \bibnamefont {Lee}},
  \bibinfo {author} {\bibfnamefont {C.}~\bibnamefont {Rockstuhl}}, \bibinfo
  {author} {\bibfnamefont {C.}~\bibnamefont {Lee}},\ and\ \bibinfo {author}
  {\bibfnamefont {K.-G.}\ \bibnamefont {Lee}},\ }\bibfield  {title} {\bibinfo
  {title} {Experimental quantum polarimetry using heralded single photons},\
  }\href@noop {} {\bibfield  {journal} {\bibinfo  {journal} {Metrologia}\
  }\textbf {\bibinfo {volume} {57}},\ \bibinfo {pages} {045008} (\bibinfo
  {year} {2020})}\BibitemShut {NoStop}%
\bibitem [{\citenamefont {Abadie}\ \emph {et~al.}(2011)\citenamefont {Abadie}
  \emph {et~al.}}]{Abadie2011}%
  \BibitemOpen
  \bibfield  {author} {\bibinfo {author} {\bibfnamefont {J.}~\bibnamefont
  {Abadie}} \emph {et~al.} (\bibinfo {collaboration} {LIGO Scientific
  Collaboration}),\ }\bibfield  {title} {\bibinfo {title} {{A gravitational
  wave observatory operating beyond the quantum shot-noise limit}},\ }\href
  {https://doi.org/10.1038/nphys2083} {\bibfield  {journal} {\bibinfo
  {journal} {Nature Physics 2011 7:12}\ }\textbf {\bibinfo {volume} {7}},\
  \bibinfo {pages} {962} (\bibinfo {year} {2011})}\BibitemShut {NoStop}%
\bibitem [{\citenamefont {Abbott}\ \emph {et~al.}(2016)\citenamefont {Abbott}
  \emph {et~al.}}]{Abbott16}%
  \BibitemOpen
  \bibfield  {author} {\bibinfo {author} {\bibfnamefont {B.~P.}\ \bibnamefont
  {Abbott}} \emph {et~al.} (\bibinfo {collaboration} {LIGO Scientific
  Collaboration and Virgo Collaboration}),\ }\bibfield  {title} {\bibinfo
  {title} {Observation of gravitational waves from a binary black hole
  merger},\ }\href {https://doi.org/10.1103/PhysRevLett.116.061102} {\bibfield
  {journal} {\bibinfo  {journal} {Phys. Rev. Lett.}\ }\textbf {\bibinfo
  {volume} {116}},\ \bibinfo {pages} {061102} (\bibinfo {year}
  {2016})}\BibitemShut {NoStop}%
\bibitem [{\citenamefont {Ghosh}\ and\ \citenamefont
  {Vitkin}(2011)}]{ghosh2011}%
  \BibitemOpen
  \bibfield  {author} {\bibinfo {author} {\bibfnamefont {N.}~\bibnamefont
  {Ghosh}}\ and\ \bibinfo {author} {\bibfnamefont {A.~I.}\ \bibnamefont
  {Vitkin}},\ }\bibfield  {title} {\bibinfo {title} {Tissue polarimetry:
  concepts, challenges, applications, and outlook},\ }\href@noop {} {\bibfield
  {journal} {\bibinfo  {journal} {Journal of biomedical optics}\ }\textbf
  {\bibinfo {volume} {16}},\ \bibinfo {pages} {110801} (\bibinfo {year}
  {2011})}\BibitemShut {NoStop}%
\bibitem [{\citenamefont {Budker}\ and\ \citenamefont
  {Romalis}(2007)}]{budker2007optical}%
  \BibitemOpen
  \bibfield  {author} {\bibinfo {author} {\bibfnamefont {D.}~\bibnamefont
  {Budker}}\ and\ \bibinfo {author} {\bibfnamefont {M.}~\bibnamefont
  {Romalis}},\ }\bibfield  {title} {\bibinfo {title} {Optical magnetometry},\
  }\href@noop {} {\bibfield  {journal} {\bibinfo  {journal} {Nature physics}\
  }\textbf {\bibinfo {volume} {3}},\ \bibinfo {pages} {227} (\bibinfo {year}
  {2007})}\BibitemShut {NoStop}%
\bibitem [{\citenamefont {Knill}\ \emph {et~al.}(2008)\citenamefont {Knill},
  \citenamefont {Leibfried}, \citenamefont {Reichle}, \citenamefont {Britton},
  \citenamefont {Blakestad}, \citenamefont {Jost}, \citenamefont {Langer},
  \citenamefont {Ozeri}, \citenamefont {Seidelin},\ and\ \citenamefont
  {Wineland}}]{knill2008randomized}%
  \BibitemOpen
  \bibfield  {author} {\bibinfo {author} {\bibfnamefont {E.}~\bibnamefont
  {Knill}}, \bibinfo {author} {\bibfnamefont {D.}~\bibnamefont {Leibfried}},
  \bibinfo {author} {\bibfnamefont {R.}~\bibnamefont {Reichle}}, \bibinfo
  {author} {\bibfnamefont {J.}~\bibnamefont {Britton}}, \bibinfo {author}
  {\bibfnamefont {R.~B.}\ \bibnamefont {Blakestad}}, \bibinfo {author}
  {\bibfnamefont {J.~D.}\ \bibnamefont {Jost}}, \bibinfo {author}
  {\bibfnamefont {C.}~\bibnamefont {Langer}}, \bibinfo {author} {\bibfnamefont
  {R.}~\bibnamefont {Ozeri}}, \bibinfo {author} {\bibfnamefont
  {S.}~\bibnamefont {Seidelin}},\ and\ \bibinfo {author} {\bibfnamefont
  {D.~J.}\ \bibnamefont {Wineland}},\ }\bibfield  {title} {\bibinfo {title}
  {Randomized benchmarking of quantum gates},\ }\href@noop {} {\bibfield
  {journal} {\bibinfo  {journal} {Physical Review A}\ }\textbf {\bibinfo
  {volume} {77}},\ \bibinfo {pages} {012307} (\bibinfo {year}
  {2008})}\BibitemShut {NoStop}%
\bibitem [{\citenamefont {Dob{\v{s}}{\'\i}{\v{c}}ek}\ \emph
  {et~al.}(2007)\citenamefont {Dob{\v{s}}{\'\i}{\v{c}}ek}, \citenamefont
  {Johansson}, \citenamefont {Shumeiko},\ and\ \citenamefont
  {Wendin}}]{dobvsivcek2007arbitrary}%
  \BibitemOpen
  \bibfield  {author} {\bibinfo {author} {\bibfnamefont {M.}~\bibnamefont
  {Dob{\v{s}}{\'\i}{\v{c}}ek}}, \bibinfo {author} {\bibfnamefont
  {G.}~\bibnamefont {Johansson}}, \bibinfo {author} {\bibfnamefont
  {V.}~\bibnamefont {Shumeiko}},\ and\ \bibinfo {author} {\bibfnamefont
  {G.}~\bibnamefont {Wendin}},\ }\bibfield  {title} {\bibinfo {title}
  {Arbitrary accuracy iterative quantum phase estimation algorithm using a
  single ancillary qubit: A two-qubit benchmark},\ }\href@noop {} {\bibfield
  {journal} {\bibinfo  {journal} {Physical Review A}\ }\textbf {\bibinfo
  {volume} {76}},\ \bibinfo {pages} {030306} (\bibinfo {year}
  {2007})}\BibitemShut {NoStop}%
\bibitem [{\citenamefont {Cram{\'e}r}(2016)}]{Cramer16}%
  \BibitemOpen
  \bibfield  {author} {\bibinfo {author} {\bibfnamefont {H.}~\bibnamefont
  {Cram{\'e}r}},\ }\href@noop {} {\emph {\bibinfo {title} {Mathematical methods
  of statistics (PMS-9)}}},\ Vol.~\bibinfo {volume} {9}\ (\bibinfo  {publisher}
  {Princeton University Press},\ \bibinfo {year} {2016})\BibitemShut {NoStop}%
\bibitem [{\citenamefont {Rao}(1992)}]{Rao92}%
  \BibitemOpen
  \bibfield  {author} {\bibinfo {author} {\bibfnamefont {C.~R.}\ \bibnamefont
  {Rao}},\ }\bibfield  {title} {\bibinfo {title} {Information and the accuracy
  attainable in the estimation of statistical parameters},\ }in\ \href@noop {}
  {\emph {\bibinfo {booktitle} {Breakthroughs in statistics}}}\ (\bibinfo
  {publisher} {Springer},\ \bibinfo {year} {1992})\ pp.\ \bibinfo {pages}
  {235--247}\BibitemShut {NoStop}%
\bibitem [{\citenamefont {Arvidsson-Shukur}\ \emph {et~al.}(2020)\citenamefont
  {Arvidsson-Shukur}, \citenamefont {Yunger~Halpern}, \citenamefont {Lepage},
  \citenamefont {Lasek}, \citenamefont {Barnes},\ and\ \citenamefont
  {Lloyd}}]{DRMAS_20_Quantum}%
  \BibitemOpen
  \bibfield  {author} {\bibinfo {author} {\bibfnamefont {D.~R.~M.}\
  \bibnamefont {Arvidsson-Shukur}}, \bibinfo {author} {\bibfnamefont
  {N.}~\bibnamefont {Yunger~Halpern}}, \bibinfo {author} {\bibfnamefont
  {H.~V.}\ \bibnamefont {Lepage}}, \bibinfo {author} {\bibfnamefont {A.~A.}\
  \bibnamefont {Lasek}}, \bibinfo {author} {\bibfnamefont {C.~H.~W.}\
  \bibnamefont {Barnes}},\ and\ \bibinfo {author} {\bibfnamefont
  {S.}~\bibnamefont {Lloyd}},\ }\bibfield  {title} {\bibinfo {title} {Quantum
  advantage in postselected metrology},\ }\href
  {https://doi.org/10.1038/s41467-020-17559-w} {\bibfield  {journal} {\bibinfo
  {journal} {Nature Communications}\ }\textbf {\bibinfo {volume} {11}},\
  \bibinfo {pages} {3775} (\bibinfo {year} {2020})}\BibitemShut {NoStop}%
\bibitem [{\citenamefont {Combes}\ \emph {et~al.}(2014)\citenamefont {Combes},
  \citenamefont {Ferrie}, \citenamefont {Jiang},\ and\ \citenamefont
  {Caves}}]{Combes14}%
  \BibitemOpen
  \bibfield  {author} {\bibinfo {author} {\bibfnamefont {J.}~\bibnamefont
  {Combes}}, \bibinfo {author} {\bibfnamefont {C.}~\bibnamefont {Ferrie}},
  \bibinfo {author} {\bibfnamefont {Z.}~\bibnamefont {Jiang}},\ and\ \bibinfo
  {author} {\bibfnamefont {C.~M.}\ \bibnamefont {Caves}},\ }\bibfield  {title}
  {\bibinfo {title} {Quantum limits on postselected, probabilistic quantum
  metrology},\ }\href {https://doi.org/10.1103/PhysRevA.89.052117} {\bibfield
  {journal} {\bibinfo  {journal} {Phys. Rev. A}\ }\textbf {\bibinfo {volume}
  {89}},\ \bibinfo {pages} {052117} (\bibinfo {year} {2014})}\BibitemShut
  {NoStop}%
\bibitem [{\citenamefont {Ferrie}\ and\ \citenamefont
  {Combes}(2014)}]{Ferrie14-2}%
  \BibitemOpen
  \bibfield  {author} {\bibinfo {author} {\bibfnamefont {C.}~\bibnamefont
  {Ferrie}}\ and\ \bibinfo {author} {\bibfnamefont {J.}~\bibnamefont
  {Combes}},\ }\bibfield  {title} {\bibinfo {title} {Weak value amplification
  is suboptimal for estimation and detection},\ }\href
  {https://doi.org/10.1103/PhysRevLett.112.040406} {\bibfield  {journal}
  {\bibinfo  {journal} {Phys. Rev. Lett.}\ }\textbf {\bibinfo {volume} {112}},\
  \bibinfo {pages} {040406} (\bibinfo {year} {2014})}\BibitemShut {NoStop}%
\bibitem [{\citenamefont {Harris}\ \emph {et~al.}(2017)\citenamefont {Harris},
  \citenamefont {Boyd},\ and\ \citenamefont {Lundeen}}]{Harris17}%
  \BibitemOpen
  \bibfield  {author} {\bibinfo {author} {\bibfnamefont {J.}~\bibnamefont
  {Harris}}, \bibinfo {author} {\bibfnamefont {R.~W.}\ \bibnamefont {Boyd}},\
  and\ \bibinfo {author} {\bibfnamefont {J.~S.}\ \bibnamefont {Lundeen}},\
  }\bibfield  {title} {\bibinfo {title} {Weak value amplification can
  outperform conventional measurement in the presence of detector saturation},\
  }\href@noop {} {\bibfield  {journal} {\bibinfo  {journal} {Phys. Rev. Lett.}\
  }\textbf {\bibinfo {volume} {118}},\ \bibinfo {pages} {070802} (\bibinfo
  {year} {2017})}\BibitemShut {NoStop}%
\bibitem [{\citenamefont {Sinclair}\ \emph {et~al.}(2017)\citenamefont
  {Sinclair}, \citenamefont {Hallaji}, \citenamefont {Steinberg}, \citenamefont
  {Tollaksen},\ and\ \citenamefont {Jordan}}]{Sinclair17}%
  \BibitemOpen
  \bibfield  {author} {\bibinfo {author} {\bibfnamefont {J.}~\bibnamefont
  {Sinclair}}, \bibinfo {author} {\bibfnamefont {M.}~\bibnamefont {Hallaji}},
  \bibinfo {author} {\bibfnamefont {A.~M.}\ \bibnamefont {Steinberg}}, \bibinfo
  {author} {\bibfnamefont {J.}~\bibnamefont {Tollaksen}},\ and\ \bibinfo
  {author} {\bibfnamefont {A.~N.}\ \bibnamefont {Jordan}},\ }\bibfield  {title}
  {\bibinfo {title} {Weak-value amplification and optimal parameter estimation
  in the presence of correlated noise},\ }\href
  {https://doi.org/10.1103/PhysRevA.96.052128} {\bibfield  {journal} {\bibinfo
  {journal} {Phys. Rev. A}\ }\textbf {\bibinfo {volume} {96}},\ \bibinfo
  {pages} {052128} (\bibinfo {year} {2017})}\BibitemShut {NoStop}%
\bibitem [{\citenamefont {Hallaji}\ \emph {et~al.}(2017)\citenamefont
  {Hallaji}, \citenamefont {Feizpour}, \citenamefont {Dmochowski},
  \citenamefont {Sinclair},\ and\ \citenamefont {Steinberg}}]{Hallaji_17_Weak}%
  \BibitemOpen
  \bibfield  {author} {\bibinfo {author} {\bibfnamefont {M.}~\bibnamefont
  {Hallaji}}, \bibinfo {author} {\bibfnamefont {A.}~\bibnamefont {Feizpour}},
  \bibinfo {author} {\bibfnamefont {G.}~\bibnamefont {Dmochowski}}, \bibinfo
  {author} {\bibfnamefont {J.}~\bibnamefont {Sinclair}},\ and\ \bibinfo
  {author} {\bibfnamefont {A.~M.}\ \bibnamefont {Steinberg}},\ }\bibfield
  {title} {\bibinfo {title} {Weak-value amplification of the nonlinear effect
  of a single photon},\ }\href {https://doi.org/10.1038/nphys4040} {\bibfield
  {journal} {\bibinfo  {journal} {Nature Physics}\ }\textbf {\bibinfo {volume}
  {13}},\ \bibinfo {pages} {540} (\bibinfo {year} {2017})}\BibitemShut
  {NoStop}%
\bibitem [{\citenamefont {Viza}\ \emph {et~al.}(2015)\citenamefont {Viza},
  \citenamefont {Mart\'{\i}nez-Rinc\'on}, \citenamefont {Alves}, \citenamefont
  {Jordan},\ and\ \citenamefont {Howell}}]{Viza15}%
  \BibitemOpen
  \bibfield  {author} {\bibinfo {author} {\bibfnamefont {G.~I.}\ \bibnamefont
  {Viza}}, \bibinfo {author} {\bibfnamefont {J.}~\bibnamefont
  {Mart\'{\i}nez-Rinc\'on}}, \bibinfo {author} {\bibfnamefont {G.~B.}\
  \bibnamefont {Alves}}, \bibinfo {author} {\bibfnamefont {A.~N.}\ \bibnamefont
  {Jordan}},\ and\ \bibinfo {author} {\bibfnamefont {J.~C.}\ \bibnamefont
  {Howell}},\ }\bibfield  {title} {\bibinfo {title} {Experimentally quantifying
  the advantages of weak-value-based metrology},\ }\href
  {https://doi.org/10.1103/PhysRevA.92.032127} {\bibfield  {journal} {\bibinfo
  {journal} {Phys. Rev. A}\ }\textbf {\bibinfo {volume} {92}},\ \bibinfo
  {pages} {032127} (\bibinfo {year} {2015})}\BibitemShut {NoStop}%
\bibitem [{\citenamefont {{Qiu}}\ \emph {et~al.}(2017)\citenamefont {{Qiu}},
  \citenamefont {{Xie}}, \citenamefont {{Liu}}, \citenamefont {{Luo}},
  \citenamefont {{Li}}, \citenamefont {{Zhang}},\ and\ \citenamefont
  {{Du}}}]{Qiu17}%
  \BibitemOpen
  \bibfield  {author} {\bibinfo {author} {\bibfnamefont {X.}~\bibnamefont
  {{Qiu}}}, \bibinfo {author} {\bibfnamefont {L.}~\bibnamefont {{Xie}}},
  \bibinfo {author} {\bibfnamefont {X.}~\bibnamefont {{Liu}}}, \bibinfo
  {author} {\bibfnamefont {L.}~\bibnamefont {{Luo}}}, \bibinfo {author}
  {\bibfnamefont {Z.}~\bibnamefont {{Li}}}, \bibinfo {author} {\bibfnamefont
  {Z.}~\bibnamefont {{Zhang}}},\ and\ \bibinfo {author} {\bibfnamefont
  {J.}~\bibnamefont {{Du}}},\ }\bibfield  {title} {\bibinfo {title} {{Precision
  phase estimation based on weak-value amplification}},\ }\href
  {https://doi.org/10.1063/1.4976312} {\bibfield  {journal} {\bibinfo
  {journal} {Applied Physics Letters}\ }\textbf {\bibinfo {volume} {110}},\
  \bibinfo {eid} {071105} (\bibinfo {year} {2017})}\BibitemShut {NoStop}%
\bibitem [{\citenamefont {Aharonov}\ \emph {et~al.}(1988)\citenamefont
  {Aharonov}, \citenamefont {Albert},\ and\ \citenamefont
  {Vaidman}}]{Vaidman88}%
  \BibitemOpen
  \bibfield  {author} {\bibinfo {author} {\bibfnamefont {Y.}~\bibnamefont
  {Aharonov}}, \bibinfo {author} {\bibfnamefont {D.~Z.}\ \bibnamefont
  {Albert}},\ and\ \bibinfo {author} {\bibfnamefont {L.}~\bibnamefont
  {Vaidman}},\ }\bibfield  {title} {\bibinfo {title} {How the result of a
  measurement of a component of the spin of a spin- \textit{1/2} particle can
  turn out to be 100},\ }\href {https://doi.org/10.1103/PhysRevLett.60.1351}
  {\bibfield  {journal} {\bibinfo  {journal} {Phys. Rev. Lett.}\ }\textbf
  {\bibinfo {volume} {60}},\ \bibinfo {pages} {1351} (\bibinfo {year}
  {1988})}\BibitemShut {NoStop}%
\bibitem [{\citenamefont {Duck}\ \emph {et~al.}(1989)\citenamefont {Duck},
  \citenamefont {Stevenson},\ and\ \citenamefont {Sudarshan}}]{Duck89}%
  \BibitemOpen
  \bibfield  {author} {\bibinfo {author} {\bibfnamefont {I.~M.}\ \bibnamefont
  {Duck}}, \bibinfo {author} {\bibfnamefont {P.~M.}\ \bibnamefont
  {Stevenson}},\ and\ \bibinfo {author} {\bibfnamefont {E.~C.~G.}\ \bibnamefont
  {Sudarshan}},\ }\bibfield  {title} {\bibinfo {title} {The sense in which a
  ``weak measurement'' of a spin-1/2 particle's spin component yields a value
  100},\ }\href {https://doi.org/10.1103/PhysRevD.40.2112} {\bibfield
  {journal} {\bibinfo  {journal} {Phys. Rev. D}\ }\textbf {\bibinfo {volume}
  {40}},\ \bibinfo {pages} {2112} (\bibinfo {year} {1989})}\BibitemShut
  {NoStop}%
\bibitem [{\citenamefont {Hosten}\ and\ \citenamefont
  {Kwiat}(2008)}]{Hosten08}%
  \BibitemOpen
  \bibfield  {author} {\bibinfo {author} {\bibfnamefont {O.}~\bibnamefont
  {Hosten}}\ and\ \bibinfo {author} {\bibfnamefont {P.}~\bibnamefont {Kwiat}},\
  }\bibfield  {title} {\bibinfo {title} {Observation of the spin hall effect of
  light via weak measurements},\ }\href
  {https://doi.org/10.1126/science.1152697} {\bibfield  {journal} {\bibinfo
  {journal} {Science}\ }\textbf {\bibinfo {volume} {319}},\ \bibinfo {pages}
  {787} (\bibinfo {year} {2008})}\BibitemShut {NoStop}%
\bibitem [{\citenamefont {Dixon}\ \emph {et~al.}(2009)\citenamefont {Dixon},
  \citenamefont {Starling}, \citenamefont {Jordan},\ and\ \citenamefont
  {Howell}}]{Dixon09}%
  \BibitemOpen
  \bibfield  {author} {\bibinfo {author} {\bibfnamefont {P.~B.}\ \bibnamefont
  {Dixon}}, \bibinfo {author} {\bibfnamefont {D.~J.}\ \bibnamefont {Starling}},
  \bibinfo {author} {\bibfnamefont {A.~N.}\ \bibnamefont {Jordan}},\ and\
  \bibinfo {author} {\bibfnamefont {J.~C.}\ \bibnamefont {Howell}},\ }\bibfield
   {title} {\bibinfo {title} {Ultrasensitive beam deflection measurement via
  interferometric weak value amplification},\ }\href
  {https://doi.org/10.1103/PhysRevLett.102.173601} {\bibfield  {journal}
  {\bibinfo  {journal} {Phys. Rev. Lett.}\ }\textbf {\bibinfo {volume} {102}},\
  \bibinfo {pages} {173601} (\bibinfo {year} {2009})}\BibitemShut {NoStop}%
\bibitem [{\citenamefont {Pang}\ \emph {et~al.}(2014)\citenamefont {Pang},
  \citenamefont {Dressel},\ and\ \citenamefont {Brun}}]{Pang14}%
  \BibitemOpen
  \bibfield  {author} {\bibinfo {author} {\bibfnamefont {S.}~\bibnamefont
  {Pang}}, \bibinfo {author} {\bibfnamefont {J.}~\bibnamefont {Dressel}},\ and\
  \bibinfo {author} {\bibfnamefont {T.~A.}\ \bibnamefont {Brun}},\ }\bibfield
  {title} {\bibinfo {title} {Entanglement-assisted weak value amplification},\
  }\href {https://doi.org/10.1103/PhysRevLett.113.030401} {\bibfield  {journal}
  {\bibinfo  {journal} {Phys. Rev. Lett.}\ }\textbf {\bibinfo {volume} {113}},\
  \bibinfo {pages} {030401} (\bibinfo {year} {2014})}\BibitemShut {NoStop}%
\bibitem [{\citenamefont {Pang}\ and\ \citenamefont {Brun}(2015)}]{Pang15}%
  \BibitemOpen
  \bibfield  {author} {\bibinfo {author} {\bibfnamefont {S.}~\bibnamefont
  {Pang}}\ and\ \bibinfo {author} {\bibfnamefont {T.~A.}\ \bibnamefont
  {Brun}},\ }\bibfield  {title} {\bibinfo {title} {Improving the precision of
  weak measurements by postselection measurement},\ }\href
  {https://doi.org/10.1103/PhysRevLett.115.120401} {\bibfield  {journal}
  {\bibinfo  {journal} {Phys. Rev. Lett.}\ }\textbf {\bibinfo {volume} {115}},\
  \bibinfo {pages} {120401} (\bibinfo {year} {2015})}\BibitemShut {NoStop}%
\bibitem [{\citenamefont {Jordan}\ \emph {et~al.}(2014)\citenamefont {Jordan},
  \citenamefont {Mart{\'\i}nez-Rinc{\'o}n},\ and\ \citenamefont
  {Howell}}]{Jordan14}%
  \BibitemOpen
  \bibfield  {author} {\bibinfo {author} {\bibfnamefont {A.~N.}\ \bibnamefont
  {Jordan}}, \bibinfo {author} {\bibfnamefont {J.}~\bibnamefont
  {Mart{\'\i}nez-Rinc{\'o}n}},\ and\ \bibinfo {author} {\bibfnamefont {J.~C.}\
  \bibnamefont {Howell}},\ }\bibfield  {title} {\bibinfo {title} {Technical
  advantages for weak-value amplification: when less is more},\ }\href@noop {}
  {\bibfield  {journal} {\bibinfo  {journal} {Physical Review X}\ }\textbf
  {\bibinfo {volume} {4}},\ \bibinfo {pages} {011031} (\bibinfo {year}
  {2014})}\BibitemShut {NoStop}%
\bibitem [{\citenamefont {Starling}\ \emph {et~al.}(2009)\citenamefont
  {Starling}, \citenamefont {Dixon}, \citenamefont {Jordan},\ and\
  \citenamefont {Howell}}]{Starling09}%
  \BibitemOpen
  \bibfield  {author} {\bibinfo {author} {\bibfnamefont {D.~J.}\ \bibnamefont
  {Starling}}, \bibinfo {author} {\bibfnamefont {P.~B.}\ \bibnamefont {Dixon}},
  \bibinfo {author} {\bibfnamefont {A.~N.}\ \bibnamefont {Jordan}},\ and\
  \bibinfo {author} {\bibfnamefont {J.~C.}\ \bibnamefont {Howell}},\ }\bibfield
   {title} {\bibinfo {title} {Optimizing the signal-to-noise ratio of a
  beam-deflection measurement with interferometric weak values},\ }\href
  {https://doi.org/10.1103/PhysRevA.80.041803} {\bibfield  {journal} {\bibinfo
  {journal} {Phys. Rev. A}\ }\textbf {\bibinfo {volume} {80}},\ \bibinfo
  {pages} {041803} (\bibinfo {year} {2009})}\BibitemShut {NoStop}%
\bibitem [{\citenamefont {Starling}\ \emph {et~al.}(2010)\citenamefont
  {Starling}, \citenamefont {Dixon}, \citenamefont {Jordan},\ and\
  \citenamefont {Howell}}]{Starling10}%
  \BibitemOpen
  \bibfield  {author} {\bibinfo {author} {\bibfnamefont {D.~J.}\ \bibnamefont
  {Starling}}, \bibinfo {author} {\bibfnamefont {P.~B.}\ \bibnamefont {Dixon}},
  \bibinfo {author} {\bibfnamefont {A.~N.}\ \bibnamefont {Jordan}},\ and\
  \bibinfo {author} {\bibfnamefont {J.~C.}\ \bibnamefont {Howell}},\ }\bibfield
   {title} {\bibinfo {title} {Precision frequency measurements with
  interferometric weak values},\ }\href
  {https://doi.org/10.1103/PhysRevA.82.063822} {\bibfield  {journal} {\bibinfo
  {journal} {Phys. Rev. A}\ }\textbf {\bibinfo {volume} {82}},\ \bibinfo
  {pages} {063822} (\bibinfo {year} {2010})}\BibitemShut {NoStop}%
\bibitem [{\citenamefont {Maga\~na Loaiza}\ \emph {et~al.}(2014)\citenamefont
  {Maga\~na Loaiza}, \citenamefont {Mirhosseini}, \citenamefont {Rodenburg},\
  and\ \citenamefont {Boyd}}]{Magana14}%
  \BibitemOpen
  \bibfield  {author} {\bibinfo {author} {\bibfnamefont {O.~S.}\ \bibnamefont
  {Maga\~na Loaiza}}, \bibinfo {author} {\bibfnamefont {M.}~\bibnamefont
  {Mirhosseini}}, \bibinfo {author} {\bibfnamefont {B.}~\bibnamefont
  {Rodenburg}},\ and\ \bibinfo {author} {\bibfnamefont {R.~W.}\ \bibnamefont
  {Boyd}},\ }\bibfield  {title} {\bibinfo {title} {Amplification of angular
  rotations using weak measurements},\ }\href
  {https://doi.org/10.1103/PhysRevLett.112.200401} {\bibfield  {journal}
  {\bibinfo  {journal} {Phys. Rev. Lett.}\ }\textbf {\bibinfo {volume} {112}},\
  \bibinfo {pages} {200401} (\bibinfo {year} {2014})}\BibitemShut {NoStop}%
\bibitem [{\citenamefont {Lyons}\ \emph {et~al.}(2015)\citenamefont {Lyons},
  \citenamefont {Dressel}, \citenamefont {Jordan}, \citenamefont {Howell},\
  and\ \citenamefont {Kwiat}}]{Lyons14}%
  \BibitemOpen
  \bibfield  {author} {\bibinfo {author} {\bibfnamefont {K.}~\bibnamefont
  {Lyons}}, \bibinfo {author} {\bibfnamefont {J.}~\bibnamefont {Dressel}},
  \bibinfo {author} {\bibfnamefont {A.~N.}\ \bibnamefont {Jordan}}, \bibinfo
  {author} {\bibfnamefont {J.~C.}\ \bibnamefont {Howell}},\ and\ \bibinfo
  {author} {\bibfnamefont {P.~G.}\ \bibnamefont {Kwiat}},\ }\bibfield  {title}
  {\bibinfo {title} {Power-recycled weak-value-based metrology},\ }\href
  {https://doi.org/10.1103/PhysRevLett.114.170801} {\bibfield  {journal}
  {\bibinfo  {journal} {Phys. Rev. Lett.}\ }\textbf {\bibinfo {volume} {114}},\
  \bibinfo {pages} {170801} (\bibinfo {year} {2015})}\BibitemShut {NoStop}%
\bibitem [{\citenamefont {Mart{\'\i}nez-Rinc{\'o}n}\ \emph
  {et~al.}(2017)\citenamefont {Mart{\'\i}nez-Rinc{\'o}n}, \citenamefont
  {Mullarkey}, \citenamefont {Viza}, \citenamefont {Liu},\ and\ \citenamefont
  {Howell}}]{Martinez17}%
  \BibitemOpen
  \bibfield  {author} {\bibinfo {author} {\bibfnamefont {J.}~\bibnamefont
  {Mart{\'\i}nez-Rinc{\'o}n}}, \bibinfo {author} {\bibfnamefont {C.~A.}\
  \bibnamefont {Mullarkey}}, \bibinfo {author} {\bibfnamefont {G.~I.}\
  \bibnamefont {Viza}}, \bibinfo {author} {\bibfnamefont {W.-T.}\ \bibnamefont
  {Liu}},\ and\ \bibinfo {author} {\bibfnamefont {J.~C.}\ \bibnamefont
  {Howell}},\ }\bibfield  {title} {\bibinfo {title} {Ultrasensitive inverse
  weak-value tilt meter},\ }\href@noop {} {\bibfield  {journal} {\bibinfo
  {journal} {Opt. Lett.}\ }\textbf {\bibinfo {volume} {42}},\ \bibinfo {pages}
  {2479} (\bibinfo {year} {2017})}\BibitemShut {NoStop}%
\bibitem [{\citenamefont {Egan}\ and\ \citenamefont {Stone}(2012)}]{Egan12}%
  \BibitemOpen
  \bibfield  {author} {\bibinfo {author} {\bibfnamefont {P.}~\bibnamefont
  {Egan}}\ and\ \bibinfo {author} {\bibfnamefont {J.~A.}\ \bibnamefont
  {Stone}},\ }\bibfield  {title} {\bibinfo {title} {Weak-value thermostat with
  0.2 mk precision},\ }\href@noop {} {\bibfield  {journal} {\bibinfo  {journal}
  {Opt. Lett}\ }\textbf {\bibinfo {volume} {37}},\ \bibinfo {pages} {4991}
  (\bibinfo {year} {2012})}\BibitemShut {NoStop}%
\bibitem [{\citenamefont {Hofmann}\ \emph {et~al.}(2012)\citenamefont
  {Hofmann}, \citenamefont {Goggin}, \citenamefont {Almeida},\ and\
  \citenamefont {Barbieri}}]{Hofmann12}%
  \BibitemOpen
  \bibfield  {author} {\bibinfo {author} {\bibfnamefont {H.~F.}\ \bibnamefont
  {Hofmann}}, \bibinfo {author} {\bibfnamefont {M.~E.}\ \bibnamefont {Goggin}},
  \bibinfo {author} {\bibfnamefont {M.~P.}\ \bibnamefont {Almeida}},\ and\
  \bibinfo {author} {\bibfnamefont {M.}~\bibnamefont {Barbieri}},\ }\bibfield
  {title} {\bibinfo {title} {Estimation of a quantum interaction parameter
  using weak measurements: Theory and experiment},\ }\href
  {https://doi.org/10.1103/PhysRevA.86.040102} {\bibfield  {journal} {\bibinfo
  {journal} {Phys. Rev. A}\ }\textbf {\bibinfo {volume} {86}},\ \bibinfo
  {pages} {040102} (\bibinfo {year} {2012})}\BibitemShut {NoStop}%
\bibitem [{\citenamefont {Kirkwood}(1933)}]{Kirkwood33}%
  \BibitemOpen
  \bibfield  {author} {\bibinfo {author} {\bibfnamefont {J.~G.}\ \bibnamefont
  {Kirkwood}},\ }\bibfield  {title} {\bibinfo {title} {Quantum statistics of
  almost classical assemblies},\ }\href {https://doi.org/10.1103/PhysRev.44.31}
  {\bibfield  {journal} {\bibinfo  {journal} {Phys. Rev.}\ }\textbf {\bibinfo
  {volume} {44}},\ \bibinfo {pages} {31} (\bibinfo {year} {1933})}\BibitemShut
  {NoStop}%
\bibitem [{\citenamefont {Dirac}(1945)}]{Dirac45}%
  \BibitemOpen
  \bibfield  {author} {\bibinfo {author} {\bibfnamefont {P.~A.~M.}\
  \bibnamefont {Dirac}},\ }\bibfield  {title} {\bibinfo {title} {On the analogy
  between classical and quantum mechanics},\ }\href
  {https://doi.org/10.1103/RevModPhys.17.195} {\bibfield  {journal} {\bibinfo
  {journal} {Rev. Mod. Phys.}\ }\textbf {\bibinfo {volume} {17}},\ \bibinfo
  {pages} {195} (\bibinfo {year} {1945})}\BibitemShut {NoStop}%
\bibitem [{\citenamefont {Johansen}(2007)}]{Johansen07}%
  \BibitemOpen
  \bibfield  {author} {\bibinfo {author} {\bibfnamefont {L.~M.}\ \bibnamefont
  {Johansen}},\ }\bibfield  {title} {\bibinfo {title} {Quantum theory of
  successive projective measurements},\ }\href
  {https://doi.org/10.1103/PhysRevA.76.012119} {\bibfield  {journal} {\bibinfo
  {journal} {Phys. Rev. A}\ }\textbf {\bibinfo {volume} {76}},\ \bibinfo
  {pages} {012119} (\bibinfo {year} {2007})}\BibitemShut {NoStop}%
\bibitem [{\citenamefont {Lundeen}\ \emph {et~al.}(2011)\citenamefont
  {Lundeen}, \citenamefont {Sutherland}, \citenamefont {Patel}, \citenamefont
  {Stewart},\ and\ \citenamefont {Bamber}}]{Lundeen11}%
  \BibitemOpen
  \bibfield  {author} {\bibinfo {author} {\bibfnamefont {J.~S.}\ \bibnamefont
  {Lundeen}}, \bibinfo {author} {\bibfnamefont {B.}~\bibnamefont {Sutherland}},
  \bibinfo {author} {\bibfnamefont {A.}~\bibnamefont {Patel}}, \bibinfo
  {author} {\bibfnamefont {C.}~\bibnamefont {Stewart}},\ and\ \bibinfo {author}
  {\bibfnamefont {C.}~\bibnamefont {Bamber}},\ }\bibfield  {title} {\bibinfo
  {title} {Direct measurement of the quantum wavefunction},\ }\href
  {https://doi.org/10.1038/nature10120} {\bibfield  {journal} {\bibinfo
  {journal} {Nature}\ }\textbf {\bibinfo {volume} {474}},\ \bibinfo {pages}
  {188} (\bibinfo {year} {2011})}\BibitemShut {NoStop}%
\bibitem [{\citenamefont {Lundeen}\ and\ \citenamefont
  {Bamber}(2012)}]{Lundeen12}%
  \BibitemOpen
  \bibfield  {author} {\bibinfo {author} {\bibfnamefont {J.~S.}\ \bibnamefont
  {Lundeen}}\ and\ \bibinfo {author} {\bibfnamefont {C.}~\bibnamefont
  {Bamber}},\ }\bibfield  {title} {\bibinfo {title} {Procedure for direct
  measurement of general quantum states using weak measurement},\ }\href
  {https://doi.org/10.1103/PhysRevLett.108.070402} {\bibfield  {journal}
  {\bibinfo  {journal} {Phys. Rev. Lett.}\ }\textbf {\bibinfo {volume} {108}},\
  \bibinfo {pages} {070402} (\bibinfo {year} {2012})}\BibitemShut {NoStop}%
\bibitem [{\citenamefont {Bamber}\ and\ \citenamefont
  {Lundeen}(2014)}]{Bamber14}%
  \BibitemOpen
  \bibfield  {author} {\bibinfo {author} {\bibfnamefont {C.}~\bibnamefont
  {Bamber}}\ and\ \bibinfo {author} {\bibfnamefont {J.~S.}\ \bibnamefont
  {Lundeen}},\ }\bibfield  {title} {\bibinfo {title} {Observing dirac's
  classical phase space analog to the quantum state},\ }\href
  {https://doi.org/10.1103/PhysRevLett.112.070405} {\bibfield  {journal}
  {\bibinfo  {journal} {Phys. Rev. Lett.}\ }\textbf {\bibinfo {volume} {112}},\
  \bibinfo {pages} {070405} (\bibinfo {year} {2014})}\BibitemShut {NoStop}%
\bibitem [{\citenamefont {Thekkadath}\ \emph {et~al.}(2016)\citenamefont
  {Thekkadath}, \citenamefont {Giner}, \citenamefont {Chalich}, \citenamefont
  {Horton}, \citenamefont {Banker},\ and\ \citenamefont
  {Lundeen}}]{Thekkadath16}%
  \BibitemOpen
  \bibfield  {author} {\bibinfo {author} {\bibfnamefont {G.~S.}\ \bibnamefont
  {Thekkadath}}, \bibinfo {author} {\bibfnamefont {L.}~\bibnamefont {Giner}},
  \bibinfo {author} {\bibfnamefont {Y.}~\bibnamefont {Chalich}}, \bibinfo
  {author} {\bibfnamefont {M.~J.}\ \bibnamefont {Horton}}, \bibinfo {author}
  {\bibfnamefont {J.}~\bibnamefont {Banker}},\ and\ \bibinfo {author}
  {\bibfnamefont {J.~S.}\ \bibnamefont {Lundeen}},\ }\bibfield  {title}
  {\bibinfo {title} {Direct measurement of the density matrix of a quantum
  system},\ }\href {https://doi.org/10.1103/PhysRevLett.117.120401} {\bibfield
  {journal} {\bibinfo  {journal} {Phys. Rev. Lett.}\ }\textbf {\bibinfo
  {volume} {117}},\ \bibinfo {pages} {120401} (\bibinfo {year}
  {2016})}\BibitemShut {NoStop}%
\bibitem [{\citenamefont {Yunger~Halpern}(2017)}]{NYH_17_Jaryznski}%
  \BibitemOpen
  \bibfield  {author} {\bibinfo {author} {\bibfnamefont {N.}~\bibnamefont
  {Yunger~Halpern}},\ }\bibfield  {title} {\bibinfo {title} {Jarzynski-like
  equality for the out-of-time-ordered correlator},\ }\href
  {https://doi.org/10.1103/PhysRevA.95.012120} {\bibfield  {journal} {\bibinfo
  {journal} {Phys. Rev. A}\ }\textbf {\bibinfo {volume} {95}},\ \bibinfo
  {pages} {012120} (\bibinfo {year} {2017})}\BibitemShut {NoStop}%
\bibitem [{\citenamefont {Yunger~Halpern}\ \emph {et~al.}(2018)\citenamefont
  {Yunger~Halpern}, \citenamefont {Swingle},\ and\ \citenamefont
  {Dressel}}]{NYH_18_Quasiprobability}%
  \BibitemOpen
  \bibfield  {author} {\bibinfo {author} {\bibfnamefont {N.}~\bibnamefont
  {Yunger~Halpern}}, \bibinfo {author} {\bibfnamefont {B.}~\bibnamefont
  {Swingle}},\ and\ \bibinfo {author} {\bibfnamefont {J.}~\bibnamefont
  {Dressel}},\ }\bibfield  {title} {\bibinfo {title} {Quasiprobability behind
  the out-of-time-ordered correlator},\ }\href
  {https://doi.org/10.1103/PhysRevA.97.042105} {\bibfield  {journal} {\bibinfo
  {journal} {Phys. Rev. A}\ }\textbf {\bibinfo {volume} {97}},\ \bibinfo
  {pages} {042105} (\bibinfo {year} {2018})}\BibitemShut {NoStop}%
\bibitem [{\citenamefont {{Gonz\'alez Alonso}}\ \emph
  {et~al.}(2019)\citenamefont {{Gonz\'alez Alonso}}, \citenamefont {{Yunger
  Halpern}},\ and\ \citenamefont {Dressel}}]{JRGA_19_Out}%
  \BibitemOpen
  \bibfield  {author} {\bibinfo {author} {\bibfnamefont {J.}~\bibnamefont
  {{Gonz\'alez Alonso}}}, \bibinfo {author} {\bibfnamefont {N.}~\bibnamefont
  {{Yunger Halpern}}},\ and\ \bibinfo {author} {\bibfnamefont {J.}~\bibnamefont
  {Dressel}},\ }\bibfield  {title} {\bibinfo {title}
  {Out-of-time-ordered-correlator quasiprobabilities robustly witness
  scrambling},\ }\href {https://doi.org/10.1103/PhysRevLett.122.040404}
  {\bibfield  {journal} {\bibinfo  {journal} {Phys. Rev. Lett.}\ }\textbf
  {\bibinfo {volume} {122}},\ \bibinfo {pages} {040404} (\bibinfo {year}
  {2019})}\BibitemShut {NoStop}%
\bibitem [{\citenamefont {Halpern}\ \emph {et~al.}(2019)\citenamefont
  {Halpern}, \citenamefont {Bartolotta},\ and\ \citenamefont
  {Pollack}}]{NYH_19_Entropic}%
  \BibitemOpen
  \bibfield  {author} {\bibinfo {author} {\bibfnamefont {N.~Y.}\ \bibnamefont
  {Halpern}}, \bibinfo {author} {\bibfnamefont {A.}~\bibnamefont
  {Bartolotta}},\ and\ \bibinfo {author} {\bibfnamefont {J.}~\bibnamefont
  {Pollack}},\ }\bibfield  {title} {\bibinfo {title} {Entropic uncertainty
  relations for quantum information scrambling},\ }\href
  {https://doi.org/10.1038/s42005-019-0179-8} {\bibfield  {journal} {\bibinfo
  {journal} {Communications Physics}\ }\textbf {\bibinfo {volume} {2}},\
  \bibinfo {pages} {1} (\bibinfo {year} {2019})}\BibitemShut {NoStop}%
\bibitem [{\citenamefont {Mohseninia}\ \emph {et~al.}(2019)\citenamefont
  {Mohseninia}, \citenamefont {Alonso},\ and\ \citenamefont
  {Dressel}}]{Razieh19}%
  \BibitemOpen
  \bibfield  {author} {\bibinfo {author} {\bibfnamefont {R.}~\bibnamefont
  {Mohseninia}}, \bibinfo {author} {\bibfnamefont {J.~R.~G.}\ \bibnamefont
  {Alonso}},\ and\ \bibinfo {author} {\bibfnamefont {J.}~\bibnamefont
  {Dressel}},\ }\bibfield  {title} {\bibinfo {title} {Optimizing measurement
  strengths for qubit quasiprobabilities behind out-of-time-ordered
  correlators},\ }\href {https://doi.org/10.1103/PhysRevA.100.062336}
  {\bibfield  {journal} {\bibinfo  {journal} {Phys. Rev. A}\ }\textbf {\bibinfo
  {volume} {100}},\ \bibinfo {pages} {062336} (\bibinfo {year}
  {2019})}\BibitemShut {NoStop}%
\bibitem [{\citenamefont {Steinberg}(1995)}]{Steinberg95}%
  \BibitemOpen
  \bibfield  {author} {\bibinfo {author} {\bibfnamefont {A.~M.}\ \bibnamefont
  {Steinberg}},\ }\bibfield  {title} {\bibinfo {title} {Conditional
  probabilities in quantum theory and the tunneling-time controversy},\ }\href
  {https://doi.org/10.1103/PhysRevA.52.32} {\bibfield  {journal} {\bibinfo
  {journal} {Phys. Rev. A}\ }\textbf {\bibinfo {volume} {52}},\ \bibinfo
  {pages} {32} (\bibinfo {year} {1995})}\BibitemShut {NoStop}%
\bibitem [{\citenamefont {Dressel}\ \emph {et~al.}(2014)\citenamefont
  {Dressel}, \citenamefont {Malik}, \citenamefont {Miatto}, \citenamefont
  {Jordan},\ and\ \citenamefont {Boyd}}]{Dressel14}%
  \BibitemOpen
  \bibfield  {author} {\bibinfo {author} {\bibfnamefont {J.}~\bibnamefont
  {Dressel}}, \bibinfo {author} {\bibfnamefont {M.}~\bibnamefont {Malik}},
  \bibinfo {author} {\bibfnamefont {F.~M.}\ \bibnamefont {Miatto}}, \bibinfo
  {author} {\bibfnamefont {A.~N.}\ \bibnamefont {Jordan}},\ and\ \bibinfo
  {author} {\bibfnamefont {R.~W.}\ \bibnamefont {Boyd}},\ }\bibfield  {title}
  {\bibinfo {title} {Colloquium: Understanding quantum weak values: Basics and
  applications},\ }\href {https://doi.org/10.1103/RevModPhys.86.307} {\bibfield
   {journal} {\bibinfo  {journal} {Rev. Mod. Phys.}\ }\textbf {\bibinfo
  {volume} {86}},\ \bibinfo {pages} {307} (\bibinfo {year} {2014})}\BibitemShut
  {NoStop}%
\bibitem [{\citenamefont {Pusey}(2014)}]{Pusey14}%
  \BibitemOpen
  \bibfield  {author} {\bibinfo {author} {\bibfnamefont {M.~F.}\ \bibnamefont
  {Pusey}},\ }\bibfield  {title} {\bibinfo {title} {Anomalous weak values are
  proofs of contextuality},\ }\href
  {https://doi.org/10.1103/PhysRevLett.113.200401} {\bibfield  {journal}
  {\bibinfo  {journal} {Phys. Rev. Lett.}\ }\textbf {\bibinfo {volume} {113}},\
  \bibinfo {pages} {200401} (\bibinfo {year} {2014})}\BibitemShut {NoStop}%
\bibitem [{\citenamefont {Arvidsson-Shukur}\ \emph {et~al.}(2017)\citenamefont
  {Arvidsson-Shukur}, \citenamefont {Gottfries},\ and\ \citenamefont
  {Barnes}}]{ArvShukur17-2}%
  \BibitemOpen
  \bibfield  {author} {\bibinfo {author} {\bibfnamefont {D.~R.~M.}\
  \bibnamefont {Arvidsson-Shukur}}, \bibinfo {author} {\bibfnamefont
  {A.~N.~O.}\ \bibnamefont {Gottfries}},\ and\ \bibinfo {author} {\bibfnamefont
  {C.~H.~W.}\ \bibnamefont {Barnes}},\ }\bibfield  {title} {\bibinfo {title}
  {Evaluation of counterfactuality in counterfactual communication protocols},\
  }\href {https://doi.org/10.1103/PhysRevA.96.062316} {\bibfield  {journal}
  {\bibinfo  {journal} {Phys. Rev. A}\ }\textbf {\bibinfo {volume} {96}},\
  \bibinfo {pages} {062316} (\bibinfo {year} {2017})}\BibitemShut {NoStop}%
\bibitem [{\citenamefont {Arvidsson-Shukur}\ and\ \citenamefont
  {Barnes}(2019)}]{ArvShukur19}%
  \BibitemOpen
  \bibfield  {author} {\bibinfo {author} {\bibfnamefont {D.~R.~M.}\
  \bibnamefont {Arvidsson-Shukur}}\ and\ \bibinfo {author} {\bibfnamefont
  {C.~H.~W.}\ \bibnamefont {Barnes}},\ }\bibfield  {title} {\bibinfo {title}
  {Postselection and counterfactual communication},\ }\href
  {https://doi.org/10.1103/PhysRevA.99.060102} {\bibfield  {journal} {\bibinfo
  {journal} {Phys. Rev. A}\ }\textbf {\bibinfo {volume} {99}},\ \bibinfo
  {pages} {060102} (\bibinfo {year} {2019})}\BibitemShut {NoStop}%
\bibitem [{\citenamefont {Kunjwal}\ \emph {et~al.}(2019)\citenamefont
  {Kunjwal}, \citenamefont {Lostaglio},\ and\ \citenamefont
  {Pusey}}]{Kunjwal19}%
  \BibitemOpen
  \bibfield  {author} {\bibinfo {author} {\bibfnamefont {R.}~\bibnamefont
  {Kunjwal}}, \bibinfo {author} {\bibfnamefont {M.}~\bibnamefont {Lostaglio}},\
  and\ \bibinfo {author} {\bibfnamefont {M.~F.}\ \bibnamefont {Pusey}},\
  }\bibfield  {title} {\bibinfo {title} {Anomalous weak values and
  contextuality: Robustness, tightness, and imaginary parts},\ }\href
  {https://doi.org/10.1103/PhysRevA.100.042116} {\bibfield  {journal} {\bibinfo
   {journal} {Phys. Rev. A}\ }\textbf {\bibinfo {volume} {100}},\ \bibinfo
  {pages} {042116} (\bibinfo {year} {2019})}\BibitemShut {NoStop}%
\bibitem [{\citenamefont {Jenne}\ and\ \citenamefont
  {Arvidsson-Shukur}(2021)}]{Jenne21}%
  \BibitemOpen
  \bibfield  {author} {\bibinfo {author} {\bibfnamefont {J.~H.}\ \bibnamefont
  {Jenne}}\ and\ \bibinfo {author} {\bibfnamefont {D.~R.~M.}\ \bibnamefont
  {Arvidsson-Shukur}},\ }\bibfield  {title} {\bibinfo {title} {Quantum
  learnability is arbitrarily distillable},\ }\href
  {https://arxiv.org/abs/2104.09520} {\bibfield  {journal} {\bibinfo  {journal}
  {arXiv preprint arXiv:1909.11116}\ } (\bibinfo {year} {2021})}\BibitemShut
  {NoStop}%
\bibitem [{\citenamefont {Jozsa}(2007)}]{Jozsa07}%
  \BibitemOpen
  \bibfield  {author} {\bibinfo {author} {\bibfnamefont {R.}~\bibnamefont
  {Jozsa}},\ }\bibfield  {title} {\bibinfo {title} {Complex weak values in
  quantum measurement},\ }\href {https://doi.org/10.1103/PhysRevA.76.044103}
  {\bibfield  {journal} {\bibinfo  {journal} {Phys. Rev. A}\ }\textbf {\bibinfo
  {volume} {76}},\ \bibinfo {pages} {044103} (\bibinfo {year}
  {2007})}\BibitemShut {NoStop}%
\bibitem [{\citenamefont {Hofmann}(2011)}]{Hofmann11}%
  \BibitemOpen
  \bibfield  {author} {\bibinfo {author} {\bibfnamefont {H.~F.}\ \bibnamefont
  {Hofmann}},\ }\bibfield  {title} {\bibinfo {title} {On the role of complex
  phases in the quantum statistics of weak measurements},\ }\href
  {https://doi.org/10.1088/1367-2630/13/10/103009} {\bibfield  {journal}
  {\bibinfo  {journal} {New Journal of Physics}\ }\textbf {\bibinfo {volume}
  {13}},\ \bibinfo {pages} {103009} (\bibinfo {year} {2011})}\BibitemShut
  {NoStop}%
\bibitem [{\citenamefont {Dressel}\ and\ \citenamefont
  {Jordan}(2012)}]{Dressel12}%
  \BibitemOpen
  \bibfield  {author} {\bibinfo {author} {\bibfnamefont {J.}~\bibnamefont
  {Dressel}}\ and\ \bibinfo {author} {\bibfnamefont {A.~N.}\ \bibnamefont
  {Jordan}},\ }\bibfield  {title} {\bibinfo {title} {Significance of the
  imaginary part of the weak value},\ }\href
  {https://doi.org/10.1103/PhysRevA.85.012107} {\bibfield  {journal} {\bibinfo
  {journal} {Phys. Rev. A}\ }\textbf {\bibinfo {volume} {85}},\ \bibinfo
  {pages} {012107} (\bibinfo {year} {2012})}\BibitemShut {NoStop}%
\bibitem [{\citenamefont {Monroe}\ \emph {et~al.}(2021)\citenamefont {Monroe},
  \citenamefont {Yunger~Halpern}, \citenamefont {Lee},\ and\ \citenamefont
  {Murch}}]{Monroe_21_Weak}%
  \BibitemOpen
  \bibfield  {author} {\bibinfo {author} {\bibfnamefont {J.~T.}\ \bibnamefont
  {Monroe}}, \bibinfo {author} {\bibfnamefont {N.}~\bibnamefont
  {Yunger~Halpern}}, \bibinfo {author} {\bibfnamefont {T.}~\bibnamefont
  {Lee}},\ and\ \bibinfo {author} {\bibfnamefont {K.~W.}\ \bibnamefont
  {Murch}},\ }\bibfield  {title} {\bibinfo {title} {Weak measurement of a
  superconducting qubit reconciles incompatible operators},\ }\href
  {https://doi.org/10.1103/PhysRevLett.126.100403} {\bibfield  {journal}
  {\bibinfo  {journal} {Phys. Rev. Lett.}\ }\textbf {\bibinfo {volume} {126}},\
  \bibinfo {pages} {100403} (\bibinfo {year} {2021})}\BibitemShut {NoStop}%
\bibitem [{\citenamefont
  {Allahverdyan}(2014)}]{Allahverdyan_14_Nonequilibrium}%
  \BibitemOpen
  \bibfield  {author} {\bibinfo {author} {\bibfnamefont {A.~E.}\ \bibnamefont
  {Allahverdyan}},\ }\bibfield  {title} {\bibinfo {title} {Nonequilibrium
  quantum fluctuations of work},\ }\href
  {https://doi.org/10.1103/PhysRevE.90.032137} {\bibfield  {journal} {\bibinfo
  {journal} {Phys. Rev. E}\ }\textbf {\bibinfo {volume} {90}},\ \bibinfo
  {pages} {032137} (\bibinfo {year} {2014})}\BibitemShut {NoStop}%
\bibitem [{\citenamefont {Miller}\ and\ \citenamefont
  {Anders}(2017)}]{Miller_17_Time}%
  \BibitemOpen
  \bibfield  {author} {\bibinfo {author} {\bibfnamefont {H.~J.~D.}\
  \bibnamefont {Miller}}\ and\ \bibinfo {author} {\bibfnamefont
  {J.}~\bibnamefont {Anders}},\ }\bibfield  {title} {\bibinfo {title}
  {Time-reversal symmetric work distributions for closed quantum dynamics in
  the histories framework},\ }\href {https://doi.org/10.1088/1367-2630/aa703f}
  {\bibfield  {journal} {\bibinfo  {journal} {New Journal of Physics}\ }\textbf
  {\bibinfo {volume} {19}},\ \bibinfo {pages} {062001} (\bibinfo {year}
  {2017})}\BibitemShut {NoStop}%
\bibitem [{\citenamefont {Levy}\ and\ \citenamefont
  {Lostaglio}(2019)}]{Levy19}%
  \BibitemOpen
  \bibfield  {author} {\bibinfo {author} {\bibfnamefont {A.}~\bibnamefont
  {Levy}}\ and\ \bibinfo {author} {\bibfnamefont {M.}~\bibnamefont
  {Lostaglio}},\ }\bibfield  {title} {\bibinfo {title} {A quasiprobability
  distribution for heat fluctuations in the quantum regime},\ }\href
  {https://arxiv.org/abs/1909.11116} {\bibfield  {journal} {\bibinfo  {journal}
  {arXiv preprint arXiv:1909.11116}\ } (\bibinfo {year} {2019})}\BibitemShut
  {NoStop}%
\bibitem [{\citenamefont {Lostaglio}(2020)}]{Lostaglio20}%
  \BibitemOpen
  \bibfield  {author} {\bibinfo {author} {\bibfnamefont {M.}~\bibnamefont
  {Lostaglio}},\ }\bibfield  {title} {\bibinfo {title} {Certifying quantum
  signatures in thermodynamics and metrology via contextuality of quantum
  linear response},\ }\href {https://doi.org/10.1103/PhysRevLett.125.230603}
  {\bibfield  {journal} {\bibinfo  {journal} {Phys. Rev. Lett.}\ }\textbf
  {\bibinfo {volume} {125}},\ \bibinfo {pages} {230603} (\bibinfo {year}
  {2020})}\BibitemShut {NoStop}%
\bibitem [{\citenamefont {Griffiths}(1984)}]{Griffiths84}%
  \BibitemOpen
  \bibfield  {author} {\bibinfo {author} {\bibfnamefont {R.~B.}\ \bibnamefont
  {Griffiths}},\ }\bibfield  {title} {\bibinfo {title} {Consistent histories
  and the interpretation of quantum mechanics},\ }\href
  {https://doi.org/https://doi.org/10.1007/BF01015734} {\bibfield  {journal}
  {\bibinfo  {journal} {Journal of Statistical Physics}\ }\textbf {\bibinfo
  {volume} {36}},\ \bibinfo {pages} {219} (\bibinfo {year} {1984})}\BibitemShut
  {NoStop}%
\bibitem [{\citenamefont {Goldstein}\ and\ \citenamefont
  {Page}(1995)}]{Goldstein95}%
  \BibitemOpen
  \bibfield  {author} {\bibinfo {author} {\bibfnamefont {S.}~\bibnamefont
  {Goldstein}}\ and\ \bibinfo {author} {\bibfnamefont {D.~N.}\ \bibnamefont
  {Page}},\ }\bibfield  {title} {\bibinfo {title} {Linearly positive histories:
  Probabilities for a robust family of sequences of quantum events},\ }\href
  {https://doi.org/10.1103/PhysRevLett.74.3715} {\bibfield  {journal} {\bibinfo
   {journal} {Phys. Rev. Lett.}\ }\textbf {\bibinfo {volume} {74}},\ \bibinfo
  {pages} {3715} (\bibinfo {year} {1995})}\BibitemShut {NoStop}%
\bibitem [{\citenamefont {Hartle}(2004)}]{Hartle04}%
  \BibitemOpen
  \bibfield  {author} {\bibinfo {author} {\bibfnamefont {J.~B.}\ \bibnamefont
  {Hartle}},\ }\bibfield  {title} {\bibinfo {title} {Linear positivity and
  virtual probability},\ }\href {https://doi.org/10.1103/PhysRevA.70.022104}
  {\bibfield  {journal} {\bibinfo  {journal} {Phys. Rev. A}\ }\textbf {\bibinfo
  {volume} {70}},\ \bibinfo {pages} {022104} (\bibinfo {year}
  {2004})}\BibitemShut {NoStop}%
\bibitem [{\citenamefont {Hofmann}(2012)}]{Hofmann12-2}%
  \BibitemOpen
  \bibfield  {author} {\bibinfo {author} {\bibfnamefont {H.~F.}\ \bibnamefont
  {Hofmann}},\ }\bibfield  {title} {\bibinfo {title} {Complex joint
  probabilities as expressions of reversible transformations in quantum
  mechanics},\ }\href {https://doi.org/10.1088/1367-2630/14/4/043031}
  {\bibfield  {journal} {\bibinfo  {journal} {New Journal of Physics}\ }\textbf
  {\bibinfo {volume} {14}},\ \bibinfo {pages} {043031} (\bibinfo {year}
  {2012})}\BibitemShut {NoStop}%
\bibitem [{\citenamefont {Hofmann}(2014)}]{Hofmann14}%
  \BibitemOpen
  \bibfield  {author} {\bibinfo {author} {\bibfnamefont {H.~F.}\ \bibnamefont
  {Hofmann}},\ }\bibfield  {title} {\bibinfo {title} {Derivation of quantum
  mechanics from a single fundamental modification of the relations between
  physical properties},\ }\href {https://doi.org/10.1103/PhysRevA.89.042115}
  {\bibfield  {journal} {\bibinfo  {journal} {Phys. Rev. A}\ }\textbf {\bibinfo
  {volume} {89}},\ \bibinfo {pages} {042115} (\bibinfo {year}
  {2014})}\BibitemShut {NoStop}%
\bibitem [{\citenamefont {Hofmann}(2015)}]{Hofmann15}%
  \BibitemOpen
  \bibfield  {author} {\bibinfo {author} {\bibfnamefont {H.~F.}\ \bibnamefont
  {Hofmann}},\ }\bibfield  {title} {\bibinfo {title} {Quantum paradoxes
  originating from the nonclassical statistics of physical properties related
  to each other by half-periodic transformations},\ }\href
  {https://doi.org/10.1103/PhysRevA.91.062123} {\bibfield  {journal} {\bibinfo
  {journal} {Phys. Rev. A}\ }\textbf {\bibinfo {volume} {91}},\ \bibinfo
  {pages} {062123} (\bibinfo {year} {2015})}\BibitemShut {NoStop}%
\bibitem [{\citenamefont {Hofmann}(2016)}]{Hofmann16}%
  \BibitemOpen
  \bibfield  {author} {\bibinfo {author} {\bibfnamefont {H.~F.}\ \bibnamefont
  {Hofmann}},\ }\bibfield  {title} {\bibinfo {title} {On the fundamental role
  of dynamics in quantum physics},\ }\href
  {https://doi.org/https://doi.org/10.1140/epjd/e2016-70086-8} {\bibfield
  {journal} {\bibinfo  {journal} {The European Physical Journal D}\ }\textbf
  {\bibinfo {volume} {70}},\ \bibinfo {pages} {118} (\bibinfo {year}
  {2016})}\BibitemShut {NoStop}%
\bibitem [{\citenamefont {Halliwell}(2016)}]{Halliwell16}%
  \BibitemOpen
  \bibfield  {author} {\bibinfo {author} {\bibfnamefont {J.~J.}\ \bibnamefont
  {Halliwell}},\ }\bibfield  {title} {\bibinfo {title} {Leggett-garg
  inequalities and no-signaling in time: A quasiprobability approach},\ }\href
  {https://doi.org/10.1103/PhysRevA.93.022123} {\bibfield  {journal} {\bibinfo
  {journal} {Phys. Rev. A}\ }\textbf {\bibinfo {volume} {93}},\ \bibinfo
  {pages} {022123} (\bibinfo {year} {2016})}\BibitemShut {NoStop}%
\bibitem [{\citenamefont {Stacey}(2019)}]{Stacey19}%
  \BibitemOpen
  \bibfield  {author} {\bibinfo {author} {\bibfnamefont {B.~C.}\ \bibnamefont
  {Stacey}},\ }\bibfield  {title} {\bibinfo {title} {Quantum theory as symmetry
  broken by vitality},\ }\href {https://arxiv.org/pdf/1907.02432} {\bibfield
  {journal} {\bibinfo  {journal} {arXiv preprint arXiv:1907.02432}\ } (\bibinfo
  {year} {2019})}\BibitemShut {NoStop}%
\bibitem [{\citenamefont {{Arvidsson-Shukur}}\ \emph
  {et~al.}(2021)\citenamefont {{Arvidsson-Shukur}}, \citenamefont {{Chevalier
  Drori}},\ and\ \citenamefont {{Yunger Halpern}}}]{DRMAS_21_Conditions}%
  \BibitemOpen
  \bibfield  {author} {\bibinfo {author} {\bibfnamefont {D.~R.~M.}\
  \bibnamefont {{Arvidsson-Shukur}}}, \bibinfo {author} {\bibfnamefont
  {J.}~\bibnamefont {{Chevalier Drori}}},\ and\ \bibinfo {author}
  {\bibfnamefont {N.}~\bibnamefont {{Yunger Halpern}}},\ }\bibfield  {title}
  {\bibinfo {title} {Conditions tighter than noncommutation needed for
  nonclassicality},\ }\href {https://doi.org/10.1088/1751-8121/ac0289}
  {\bibfield  {journal} {\bibinfo  {journal} {Journal of Physics A:
  Mathematical and Theoretical}\ }\textbf {\bibinfo {volume} {54}},\ \bibinfo
  {pages} {284001} (\bibinfo {year} {2021})}\BibitemShut {NoStop}%
\bibitem [{\citenamefont {{De Bievre}}(2021)}]{DeBievre_21_Kirkwood}%
  \BibitemOpen
  \bibfield  {author} {\bibinfo {author} {\bibfnamefont {S.}~\bibnamefont {{De
  Bievre}}},\ }\bibfield  {title} {\bibinfo {title} {{Kirkwood-Dirac
  nonclassicality, support uncertainty and complete incompatibility}},\
  }\href@noop {} {\bibfield  {journal} {\bibinfo  {journal} {arXiv e-prints}\
  ,\ \bibinfo {eid} {arXiv:2106.10017}} (\bibinfo {year} {2021})},\ \Eprint
  {https://arxiv.org/abs/2106.10017} {arXiv:2106.10017 [quant-ph]} \BibitemShut
  {NoStop}%
\bibitem [{\citenamefont {{Oszmaniec}}\ \emph {et~al.}(2021)\citenamefont
  {{Oszmaniec}}, \citenamefont {{Brod}},\ and\ \citenamefont
  {{Galv{\~a}o}}}]{Oszmaniec_21_Measuring}%
  \BibitemOpen
  \bibfield  {author} {\bibinfo {author} {\bibfnamefont {M.}~\bibnamefont
  {{Oszmaniec}}}, \bibinfo {author} {\bibfnamefont {D.~J.}\ \bibnamefont
  {{Brod}}},\ and\ \bibinfo {author} {\bibfnamefont {E.~F.}\ \bibnamefont
  {{Galv{\~a}o}}},\ }\bibfield  {title} {\bibinfo {title} {{Measuring
  relational information between quantum states, and applications}},\
  }\href@noop {} {\bibfield  {journal} {\bibinfo  {journal} {arXiv e-prints}\
  ,\ \bibinfo {eid} {arXiv:2109.10006}} (\bibinfo {year} {2021})},\ \Eprint
  {https://arxiv.org/abs/2109.10006} {arXiv:2109.10006 [quant-ph]} \BibitemShut
  {NoStop}%
\bibitem [{\citenamefont {Braunstein}\ and\ \citenamefont
  {Caves}(1994)}]{Braunstein94}%
  \BibitemOpen
  \bibfield  {author} {\bibinfo {author} {\bibfnamefont {S.~L.}\ \bibnamefont
  {Braunstein}}\ and\ \bibinfo {author} {\bibfnamefont {C.~M.}\ \bibnamefont
  {Caves}},\ }\bibfield  {title} {\bibinfo {title} {Statistical distance and
  the geometry of quantum states},\ }\href
  {https://doi.org/10.1103/PhysRevLett.72.3439} {\bibfield  {journal} {\bibinfo
   {journal} {Phys. Rev. Lett.}\ }\textbf {\bibinfo {volume} {72}},\ \bibinfo
  {pages} {3439} (\bibinfo {year} {1994})}\BibitemShut {NoStop}%
\bibitem [{\citenamefont {Nielsen}\ and\ \citenamefont
  {Chuang}(2011)}]{Nielsen11}%
  \BibitemOpen
  \bibfield  {author} {\bibinfo {author} {\bibfnamefont {M.~A.}\ \bibnamefont
  {Nielsen}}\ and\ \bibinfo {author} {\bibfnamefont {I.~L.}\ \bibnamefont
  {Chuang}},\ }\href@noop {} {\emph {\bibinfo {title} {Quantum Computation and
  Quantum Information: 10th Anniversary Edition}}},\ \bibinfo {edition} {10th}\
  ed.\ (\bibinfo  {publisher} {Cambridge University Press},\ \bibinfo {address}
  {New York, NY, USA},\ \bibinfo {year} {2011})\BibitemShut {NoStop}%
\bibitem [{\citenamefont {Wiseman}(2002)}]{Wiseman_02_Weak}%
  \BibitemOpen
  \bibfield  {author} {\bibinfo {author} {\bibfnamefont {H.~M.}\ \bibnamefont
  {Wiseman}},\ }\bibfield  {title} {\bibinfo {title} {Weak values, quantum
  trajectories, and the cavity-qed experiment on wave-particle correlation},\
  }\href {https://doi.org/10.1103/PhysRevA.65.032111} {\bibfield  {journal}
  {\bibinfo  {journal} {Phys. Rev. A}\ }\textbf {\bibinfo {volume} {65}},\
  \bibinfo {pages} {032111} (\bibinfo {year} {2002})}\BibitemShut {NoStop}%
\end{thebibliography}%

\end{document}